\documentclass[12pt,a4paper]{article}   
\usepackage[dvips]{graphicx}
\usepackage{amsmath,amsfonts,amssymb}  

\def\m@th{\mathsurround=0pt}

\newcommand{\pn}{\par\noindent}

\makeatletter
\renewcommand{\appendix}{
\renewcommand{\thesection}{Appendix \Alph{section}\hspace{-.5em}}
\renewcommand{\thesubsection}
 {\Alph{section}\arabic{subsection}.\hspace{-.5em}}
\renewcommand{\thesubsubsection}
 {\Alph{section}\arabic{subsection}.\arabic{subsubsection}.\hspace
                                                            {-.5em}}
\@addtoreset{equation}{subsection}
\renewcommand{\theequation}{\Alph{section}.\arabic{equation}} 
\setcounter{section}{0}}
\makeatother

\title{
Elementary functions\\ in\\Thermodynamic Bethe Ansatz\footnote{
Based on talks given at   ``Infinite Analysis 2014" (Tokyo, Feb. 2014)
and at  ``Integrable lattice models and quantum field theories" (Bad-Honnef, June 2014)}
}
\author{ J. Suzuki\thanks{e-mail: sjsuzuk@ipc.shizuoka.ac.jp}\\
        \parbox{0.9\textwidth}{
        {\em
        \begin{center}
       Department of Physics, Faculty of Science\\
       Shizuoka University,\\
      Ohya 836, Shizuoka,\\
       Japan
        \end{center}
        }}
       }
\date{January 2015}
\begin{document}
\maketitle
\begin{abstract}

Some years ago, Fendley found  an  explicit solution to the thermodynamic Bethe ansatz  (TBA)  equation
for a ${\cal N}=2$  supersymmetric  theory in 2D  with a specific F-term.
Motivated by this,   we seek for
explicit solutions   for other 
super-potential cases 
utilizing  the idea from the ODE/IM correspondence.
We find that  the TBA equations, corresponding to a wider class of super-potentials, 
 admit  solutions in terms of elementary functions
such as modified Bessel functions and confluent hyper-geometric series.
\end{abstract}
\clearpage

\section{Introduction}\label{sec1}
The \underline{T}hermodynamic \underline{B}ethe \underline{A}nsatz (TBA) is  one of the most efficient tools in
the field of integrable systems \cite{Takahashibook}.
Once  input data such as factorized S- matrices \cite{AlZPotts, DTP}, special patterns of Bethe ansatz roots (string hypothesis) \cite{Gaudin, TakahahiSuzuki},
or the fusion relations \cite{KlumperRSOS, JKSfusion} are given,  it provides finitely or infinitely many 
coupled integrals equations as output.
These equations make the quantitative analyses possible in 
integrable  1+1D quantum field theories  of finite size \cite{AlZPotts} or 
1D quantum systems at finite temperatures \cite{Takahashibook}.
The numerical analysis provides  physical quantities such as 
the specific heat or the magnetic susceptibility for the whole range of 
temperature \cite{TsvelikWiegmann}, or  flow of the $g$ function by the change in the system size \cite{DRTW}.
On the other hand, some limited information, 
such as the central charge,  is available analytically from the TBA equations.
This is due to the fact that the nonlinearity of the TBA equations defies  explicit solutions
in most of cases.

Some years ago, Fendley \cite{Fendley99}  obtained a rare example: an explicit solution in  the
massless limit of an  integrable 
${\cal N}=2$  supersymmetric theory in 2D. 
%
There is a deep structure behind the model  which
connects the solution to the (massive) TBA equation 
and the solution to  the Painlev{\'e} III  (PIII) equation.
The proof  in \cite{Fendley99}  replies on the heavy machinery on the 
solution to the PIII equation which has been developed  in \cite{MTW, AlZ, TW}.
Especially it utilizes the Tracy-Widom representation to the PIII  solution,
  valid for a massive theory in general, while the explicit solution
in terms of  elementary function is possible only in the massless case.
Then one may wonder if any simpler derivation is possible for the
result in \cite{Fendley99}, as the massless theory possesses a larger symmetry and thus 
offers a  simpler structure.
 
In this communication, we will argue that  the \underline{O}rdinary  \underline{D}ifferential  
\underline{E}quation/ \underline{I}ntegrable  \underline{M}odel (ODE/IM)
\footnote{For review see \cite{DDT}}
 correspondence provides a much simpler
explanation of the solution.  
This program was actually suggested in \cite{Fendley99}.
We will make it concrete.
The Stokes multipliers  $\tau$,  associated with a  simple ODE with an irregular singularity at infinity, 
turns out to  provide the solution in \cite{Fendley99}. 
This  may sound odd as there seems to be no relation between the original problem
and the ODE thus there is no reason to consider a specific ODE.
There is, however,  a relation.  
We take the super-potential corresponding to Fendley's solution. 
From the  potential, we construct a  function, which solves an ODE.
At this stage,  the Stokes multiplier is a trivial constant.  We then ``deform"  the ODE by 
a weak gauge field (or  small angular momentum).  
Remarkably, the first nontrivial response of the Stokes multiplier
to the  gauge field reproduces the solution in \cite{Fendley99}.

This immediately leads to a generalization. 
There exists a list of relevant super-potentials \cite{VafaWarner, Martinec, CV} for the Landau-Ginzburg
description of super-conformal theories.  
The corresponding TBA equations for 
perturbed cases are partially derived  in \cite{FI, CFIV}.
 Thus, starting from  one of the available potentials,
we can construct an ODE  and evaluate the first non trivial  response to the weak gauge field.
The resultant Stokes multipliers are then transformed automatically into Y functions.
We will show that these Y functions solve TBA equations  in perturbed ${\cal N}=2$ minimal  theories
with the  SU(2)$_k$ and with the SU(3)$_1$ chiral rings,
 which generalize Fendley's solution for the SU(2)$_1$.

This paper is organized as follows.
Section \ref{review-ode}  is devoted  to a short review on the ODE/IM correspondence.
 Fendley's solution is re-derived from the Stokes multiplier associated to a special ODE in section \ref{Fendley}.
In section \ref{sec:generalize}, we apply the working hypothesis obtained in the previous section 
to  the TBA equations for  perturbed ${\cal N}=2$  minimal
theories in 2D  with the SU(2)$_k$ and the SU(n)$_1$  chiral rings. 
We demonstrate the  applications of the exact solutions  in  section 
\ref{sec:application}. 
Section \ref{sec:summary} is devoted to a summary and  future problems.
\section{The ODE/IM correspondence}\label{review-ode}

We  summarize  results from  the ODE/IM correspondence which are relevant in the following discussions.
For details, see \cite{DDT}.

We consider  a simple ODE of {$n^\text{th}$} order 
in the complex plane $x \in \mathbb{C}$,
\begin{equation}
 \Bigl( (-)^{n-1}\frac{d^n}{dx^n} + (x^{n \alpha} -E) \Bigr ) \psi(x,E)  = 0,
\label{j:Schr}
\end{equation}
where  $\alpha \in \mathbb{R}_{\ge -1}$.

Since it has the irregular singularity at $\infty$, 
we conveniently divide the complex plane into sectors.
Let ${\mathcal S}_j$ be a sector in the complex plane, 
$$
{\mathcal S}_j=\Bigl\{ x \Big\vert \;
\big\vert{\rm arg}\, x - \frac{2j \pi}{n(\alpha+1)} \big\vert 
< \frac{\pi}{n(\alpha+1)}
\Bigr\}.
$$
The sector ${\mathcal S}_0$ thus includes the positive real axis.

Let $\phi(x, E)$ be a solution to (\ref{j:Schr}) 
which decays exponentially 
as $x$ tends to $\infty$  inside ${\mathcal S}_0$,
\begin{equation}
\frac{d^p \phi(x,E) }{dx^p} \sim (-1)^p \frac{ x^{(1-n+2p)\frac{\alpha}{2}} }{\sqrt{n} i^{ (n-1)/2} }
\exp\Bigl(-\frac{ x^{\alpha+1} }{\alpha+1}\Bigr), 
\qquad x \in {\mathcal S}_0
\label{j:asym}   
\end{equation}
for $p\in \mathbb{Z}_{\ge 0}$.

The crucial observation in \cite{DT99} is the
``discrete rotational symmetry" of
 (\ref{j:Schr}):  the invariance under the simultaneous transformations
\begin{align}
x &\rightarrow q^{-1} x,&     E & \rightarrow  E \Omega^n,   \nonumber \\
&q={\rm e}^{\frac{2\pi i}{n(\alpha+1)}},&    &\Omega=q^{-\alpha}.   \label{rotationsymm} 
\end{align}
  We then introduce  
  \begin{equation}\label{yjsu2}
  \phi_j= q^{(n-1)j/2} \phi(q^{-j} x,  \Omega^{n j} E).
 \end{equation}
Thanks to the  discrete rotational symmetry, 
any $ \phi_j\,(j \in \mathbb{Z})$ is  a solution to  (\ref{j:Schr}).  
The set 
$( \phi_j, \cdots,  \phi_{j+n-1})$ forms the \underline{F}undamental \underline{S}ystem of 
\underline{S}olutions (FSS) 
in ${\mathcal S}_j$.
To see this, we introduce
the Wronskian matrix $\Phi^{(m)}_{j_1,\cdots,j_m}(x,E)$ and the Wronskian
 $W^{(m)}_{j_1,\cdots,j_m}(x,E)={\rm det}\Phi^{(m)}_{j_1,\cdots,j_m}(x,E) $,
\[
\Phi^{(m)}_{j_1,\cdots,j_m}(x,E)=
\begin{pmatrix}
 \phi_{j_1} &    \cdots  &   \phi_{j_m}\\
\vdots&             &     \vdots \\
 \phi^{(m-1)}_{j_1} &    \cdots  &   \phi^{(m-1)}_{j_m}\\
\end{pmatrix}
\]
where $m\le n$.
Especially, when suffixes $\{j\}$ are  consecutive integers, e.g., $j_k=j+k-1$
we write simply  $\Phi^{(m)}_j(x,E)$ and their determinants  $W^{(m)}_j(x,E)$ 
(we drop the $x$ dependency when $m=n$).
By using the asymptotic form (\ref{j:asym}),  
one can check $W^{(n)}_j(E)=1$, hence 
the set $( \phi_j, \cdots,  \phi_{j+n-1})$ is linearly independent.

We are interested in the relation among the FSS in different sectors.
Let us start from  the relation between ${\mathcal S}_0$ and   ${\mathcal S}_1$.
Two Wronskian matrices $\Phi^{(n)}_0$ and $\Phi^{(n)}_1$ are simply connected by
\begin{equation}\label{ODEconnection}
\Phi^{(n)}_{0} =\Phi^{(n)}_{1}{\cal M}^{(n)}(E) 
\end{equation}
where
\begin{equation}\label{Stokesmatrix}
{\cal M}^{(n)} (E)=
\begin{pmatrix}
\tau^{(1)}_1(E)&   1&   0& \cdots& 0  \\
-\tau^{(2)}_1(E)&   0&  1& \cdots& 0  \\ 
\vdots&                &    &           &     1\\
(-1)^{n-1}\tau^{(n)}_1(E)&   0&  0& \cdots& 0\\
\end{pmatrix}.
\end{equation}
The entries $\tau^{(a)}_1(E)$ are called the Stokes multipliers.
The discrete rotational symmetry then results in
\begin{equation}
\Phi^{(n)}_{j} =\Phi^{(n)}_{j+1}{\cal M}^{(n)}_1(E \Omega^{nj}) .
\end{equation}
The first observation of the ODE/IM correspondence is that 
this linear relation, evaluated at the origin, can be identified with 
Baxter's TQ relation \cite{Baxterbook} for $n=2$.
That is, let 
\begin{equation} \label{defQpmleq0}
{\mathbf T}_1(E)=\tau^{(1)}_1(E \Omega^{-2}), \qquad   Q^-(E)= \phi(0, E),  \qquad Q^+(E)= \phi'(0, E).
\end{equation}
Then one  presents   $\tau_1 =  \phi_0 / \phi_1+  \phi_2/ \phi_1$ equivalently as
\begin{equation}\label{DVF1}
{\mathbf T}_1(E) = q^{\mp\frac{1}{2}}\frac{ Q^{\mp} (E \Omega^{-2}) }{Q^{\mp}(E)}  +  q^{\pm\frac{1}{2}} \frac{Q^{\mp}(E \Omega^2)}{ Q^{\mp}(E)} .
\end{equation}
This is known as the \underline{D}ressed \underline{V}acuum \underline{F}orm (DVF) in integrable models.

Suppose that wave functions  are given in advance. 
Then,  thanks to the normalization of $\phi$, the Stokes multipliers are represented by wave functions, e.g., 
\begin{equation} \label{qWronskian0}
\tau^{(1)}_1(E) =W^{(n)}_{0,2,\cdots,n}(E).
\end{equation}
Let us introduce more generally 
\begin{equation}\label{deftau}
\tau^{(a)}_m (E)=W^{(n)}_{0,\cdots, a-1, a+m,\cdots,n+m-1}(E).
\end{equation}
Some of them appear in  the connection problem between
${\cal S}_0$ and  ${\cal S}_m$ \cite{JSsuN}.

The identity among Wronskians implies
\begin{equation}
\tau^{(a)}_{m}(E) \tau^{(a)}_m (E\Omega^n) = \tau^{(a+1)}_{m}(E)    \tau^{(a-1)}_{m}(E \Omega^n)  + 
 \tau^{(a)}_{m+1}(E)    \tau^{(a)}_{m-1}(E \Omega^n)
\label{j:tauT}
\end{equation}
where $1\le a \le n-1$, $m\in \mathbb{Z}_{\ge 1}$,
$\tau^{(0)}_m=\tau^{(n)}_m=1$ and $\tau^{(a)}_0=1$.

After a suitable shift of the parameters and a change of variables 
($E \rightarrow v, \tau \rightarrow T$), 
one arrives at  the SU(n) T system \cite{KNS}, 
\begin{equation}
{\mathbf T}^{(a)}_m(v+i){\mathbf T}^{(a)}_m(v-i)={\mathbf T}^{(a+1)}_m(v) {\mathbf T}^{(a-1)}_m(v) +{\mathbf T}^{(a)}_{m+1}(v)  {\mathbf T}^{(a)}_{m-1}(v).
\label{j:Tode}
\end{equation}
The conditions  $ {\mathbf T}^{(a)}_{0}(v)={\mathbf T}^{(0)}_{m}(v)={\mathbf T}^{(n)}_{m}(v)=1$ are again imposed.   
By employing the further  transformation \cite{KluemperPearce, KNS},
\begin{equation}\label{YTtransformation}
Y^{(a)}_m(v)= \frac{{\mathbf T}^{(a)}_{m-1}(v) {\mathbf T}^{(a)}_{m+1}(v)}{   {\mathbf T}^{(a+1)}_m(v) {\mathbf T}^{(a-1)}_m(v)},
\end{equation}
one obtains the SU(n) Y system,
\begin{equation}
Y^{(a)}_m(v+i) Y^{(a)}_m(v-i) =\frac{ (1+Y^{(a)}_{m-1}(v))(1+Y^{(a)}_{m+1}(v))}{   (1+(Y^{(a+1)}_{m}(v))^{-1})(1+(Y^{(a-1)}_{m}(v))^{-1})     }   .
\end{equation}
The Y system for ADE  scattering models was originally introduced in \cite{ZamolodchikovY}. 

It is well known under the assumption of the analytic properties on $Y^{(a)}_m$ that 
the above algebraic equations can be transformed into the TBA equations.
This also manifests the  ODE/IM correspondence.

We can also formulate the problem on the positive real axis.
To simplify notations, let us concentrate on the case $n=2$ (the radial Schr{\"o}dinger problem),
\begin{equation}
 \Bigl( -\frac{d^2}{dx^2} + (x^{2 \alpha} -E)+\frac{\ell(\ell+1)}{x^2}  \Bigr ) \psi(x,E,\ell)  = 0.
\label{raialSchr}
\end{equation}
This is also regarded as the introduction of a gauge field when rewriting it as
\begin{equation}
 \Bigl( -(\frac{d}{dx}-\frac{\ell}{x})(\frac{d}{dx}+\frac{\ell}{x})  + (x^{2 \alpha} -E)
\Bigr ) \psi(x,E,\ell)  = 0.
\label{raialSchrII}
\end{equation}

While, in  the absence of the gauge field, the Q function is directly related to the
value of the wave function at the origin, as  in (\ref{defQpmleq0}), 
it is no longer the case  with the presence of  the gauge field.
It is however shown in \cite{BLZODE, DT99} that the Q function appears naturally
if one considers the connection problem of the FSS near the origin and the FSS
at large $x$.
Denote  two solutions near the origin, 
\[
\chi^{\pm}(x,E,\ell)  \sim  \frac{1}{\sqrt{2\ell+1}}   x^{ \pm (\ell+\frac{1}{2}) +\frac{1}{2} }
\]
and   set more generally, analogously to (\ref{yjsu2}),
\[
\chi^{\pm}_j (x,E,\ell) =q^{\frac{j}{2}} \chi^{\pm}(q^{-j} x, \Omega^{2j} E, \ell).
\]
The $x \rightarrow 0$ behavior implies
\[
\chi^{\pm}_j (x,E,\ell) =q^{\mp j(\ell+\frac{1}{2})} \chi^{\pm}(x, E, \ell).
\]

The ``radial" connection relation is given by  \footnote{We change the sign of $D^+$ from \cite{DT99}} ,
\begin{equation}\label{radialconnection}
 \phi(x,E,\ell) =\mathbf{ D}^{-}(E,\ell) \chi^-(x,E,\ell) + \mathbf{ D}^{+}(E,\ell) \chi^+(x,E,\ell), 
\end{equation}
or equivalently,
\[
 \phi_j(x,E,\ell) = \mathbf{ D}^{-}(E \Omega^{2j} ,\ell) \chi^-_j(x,E,\ell) + \mathbf{ D}^{+}(E \Omega^{2j},\ell) \chi^+_j(x,E,\ell).
\]
The connection relations among $ \phi_j(x,E,\ell)$ assume the same form, e.g,  (\ref{ODEconnection}).
One then derives the DVF in the radial problem as
\begin{equation}\label{DVF2}
{\mathbf T}_1(E,\ell) = q^{\mp(\frac{1}{2}+\ell) }\frac{ \mathbf{ D}^{\mp} (E \Omega^{-2},\ell) }{\mathbf{ D}^{\mp}(E,\ell)}  +  q^{\pm(\frac{1}{2}+\ell)} \frac{\mathbf{ D}^{\mp}(E \Omega^2,\ell)}{ \mathbf{ D}^{\mp}(E,\ell)} .
\end{equation}
By comparing with  (\ref{DVF1}), one concludes that $\mathbf{ D}^{\pm}$ generalizes $Q^{\pm}$ for non-zero $\ell$ case.
They recover (\ref{DVF1}) by putting $\ell=0$.
In terms of  $\mathbf{ D}^{\pm}$, 
the Wronskian representation of the generalized Stokes multipliers  (\ref{qWronskian0}) is given by
\begin{align}\label{qWronskian1}
{\mathbf T}_j(E,\ell)=&q^{-(j+1)(\ell+\frac{1}{2})}  \mathbf{ D}^{+}(E \Omega^{j+1},\ell) \mathbf{ D}^{-}(E \Omega^{-(j+1)},\ell)  \nonumber \\
&-q^{(j+1)(\ell+\frac{1}{2})}  \mathbf{ D}^{-}(E \Omega^{j+1},\ell) \mathbf{ D}^{+}(E \Omega^{-(j+1)},\ell)  .
\end{align}

This is to be identified with the quantum Wronskian relation \cite{BLZ2},
except for  a difference in normalization as discussed in \cite{DT99}.

\section{Revisiting Fendley's solution}\label{Fendley}

In \cite{FI, CFIV}, a class of integrable  ${\cal N}$=2 supersymmetric theories in 2D,
 described by  Landau-Ginzburg actions, has been analyzed. 
For models with spontaneously broken ${\mathbf Z}_n$ symmetry
 a set of TBA equations has been proposed. 
 Especially in the latter paper, direct relations of the solution to TBA equations and 
solutions to PIII  or to affine Toda equations are
argued.

When the super-potential is given by  $W(X)= \frac{X^3}{3}-X$,
the explicit TBA equations read,
\begin{align}
A(\theta, \mu) &=2 u(\theta, \mu) -\int_{-\infty}^{\infty} 
\frac{d\theta'}{2\pi} \frac{1}{\cosh(\theta-\theta') }\ln(1+ B(\theta', \mu)^2),  \nonumber \\
B(\theta, \mu) &= -\int_{-\infty}^{\infty} \frac{d\theta'}{2\pi} \frac{1}{\cosh(\theta-\theta')} {\rm e}^{-A(\theta', \mu)}.  \label{N2TBA1}
\end{align}
In the above, $u(\theta, \mu)=  \mu\cosh \theta$  and  $ \mu$ corresponds to a physical mass.
It reduces to ${\rm e}^{\theta}/2$ in the massless limit \cite{AlZamolodchikovmassless}. 

Fendley found  the following explicit solution  \cite{Fendley99} for the massless case,
\begin{align}
{\rm e}^{-A(\theta)} &= -2\pi \frac{d}{dz} ({\rm Ai}(z))^2,   \nonumber  \\
B(\theta)  &= 2\pi \frac{d}{dz}  {\rm Ai}(z{\rm e}^{i\frac{\pi}{3}}) {\rm Ai}(z{\rm e}^{-i\frac{\pi}{3}})  \label{Fendelysolution}
\end{align}
where $z=(3 {\rm e}^{\theta}/4)^{2/3}$.

We  will re-derive the solution from the ODE side,   starting  from,
\begin{equation}\label{AiryODE}
\Bigl(- \frac{d^2}{dx^2} +(x-E) \Bigr) \psi(x,E)=0
\end{equation}
which is the case $(n,\alpha)=(2,\frac{1}{2})$ in (\ref{j:Schr}).
It follows from (\ref{rotationsymm}) that 
\[
q={\rm e}^{\frac{2\pi}{3}i}, \qquad \Omega ={\rm e}^{-\frac{\pi}{3}i} .
\]
It is well known that the Airy function  solves this equation. 
Respecting the leading asymptotic from (\ref{j:asym}), the desired solution of  (\ref{AiryODE}) is given by
\begin{equation}\label{solAiry}
\phi(x, E)= \sqrt{\frac{2\pi}{i}} {\rm Ai}(x-E).
\end{equation}

This immediately gives $Q^{\pm}$ 
\footnote{
Actually, the role played by the Airy function in ${\cal N}=2$ SUSY theory, especially its
relation to $Q^{\pm}$  has been
firstly noted  in \cite{DT98},  independently from  \cite{Fendley99}, exactly in the context of the ODE/IM  correspondence.
}
 in (\ref{defQpmleq0})
\begin{align}\label{explicitQpmk1}
\begin{cases}
Q^{-}(E) &= \sqrt{\frac{2\pi}{i}}{\rm Ai}(-E),  \\
Q^{+}(E) &=- \sqrt{\frac{2\pi}{i}}\frac{d}{dE} {\rm Ai}(-E).  \\
\end{cases}
\end{align}
We remark that the ODE (\ref{AiryODE})  is not totally independent of the original problem.

Although in  the ${\cal N}=2$ symmetric theory, the argument of the super-potential
$W(X)$ is a super-field $X$,   we allow for a usual variable in $W(x)$. 
Then the solution 
$\phi(x, E)$   has a well known  integral representation,
\begin{equation}
\phi(x,E)= \int_{\cal C} {\rm e}^{ (x-E)^{\frac{3}{2}}W(z (x-E)^{-\frac{1}{2}} ) } dz.
\label{integralrep}
\end{equation}
The  contour must be chosen so as to reproduce the 
asymptotic behavior (\ref{j:asym}) of $\phi$.
Respecting the ``discrete rotational symmetry"  and
the change of the integration contour, it is  easy to show
\begin{equation}\label{connectionAiry}
\phi(x,E)+{\rm e}^{\frac{2\pi}{3}i} \phi({\rm e}^{\frac{2\pi}{3}i} x, {\rm e}^{\frac{2\pi}{3}i} E) +
{\rm e}^{-\frac{2\pi}{3}i} \phi({\rm e}^{-\frac{2\pi}{3}i} x, {\rm e}^{-\frac{2\pi}{3}i} E) =0.
\end{equation}
This  is equivalent to
the three terms relation for the Airy function and it  leads to the conclusion $\tau_1={\mathbf T}_1=1$.
We can easily check this by using the DVF(\ref{DVF1}) and
(\ref{solAiry}).  By choosing  the upper index in (\ref{DVF1}) we have 
\[
{\mathbf T}_1(E) ={\rm e}^{-\frac{\pi}{3}i}   \frac{ {\rm Ai}(E {\rm e}^{\frac{2\pi}{3}i})}{{\rm Ai}(E)}
+ {\rm e}^{\frac{\pi}{3}i}   \frac{ {\rm Ai}(E {\rm e}^{-\frac{2\pi}{3}i})}{{\rm Ai}(E)}
=- {\rm e}^{\frac{2\pi}{3}i}   \frac{ {\rm Ai}(E {\rm e}^{\frac{2\pi}{3}i})}{{\rm Ai}(E)}
- {\rm e}^{-\frac{2\pi}{3}i}   \frac{ {\rm Ai}(E {\rm e}^{-\frac{2\pi}{3}i})}{{\rm Ai}(E)}.
\]
Thus   ${\mathbf T}_1=1$ thanks to (\ref {connectionAiry}).

Now T system is trivially represented as 
\begin{equation} \label{trivialT}
{\mathbf T}_1^2=1,  \qquad 
{\mathbf T}_2=0.
\end{equation}
It simply gives a  trivial solution of TBA, $Y_1=0$, which  is far from Fendley's solution.

We then ``deform" the  ODE  by the nonzero angular momentum term as suggested in \cite{Fendley99},
\begin{equation}\label{AiryWithL}
\Bigl(   -(\frac{d}{dx}-\frac{\ell}{x})(\frac{d}{dx}+\frac{\ell}{x}) +x-E \Bigr) \psi(x,E, \ell)=0 .
\end{equation}
%
%
Below we will argue that  this replacement leads  to the desired T, Y system and 
to  the TBA.

The  Stokes multiplier has the form  (\ref{DVF2}),
\begin{align}\label{DVF3}
{\mathbf T}_1(E,\ell) &=\xi^{\mp}\frac{ \mathbf{ D}^{\mp} (E \Omega^{-2},\ell) }{\mathbf{ D}^{\mp}(E,\ell)}  +  \xi^{\pm}
 \frac{\mathbf{ D}^{\mp}(E \Omega^2,\ell)}{ \mathbf{ D}^{\mp}(E,\ell)},   \\
& \xi = {\rm e}^{\frac{\pi - h}{3}i}  \label{xikeq1}
\end{align}
where $h=-2\ell\pi$. 
 We assume that the following  limit exists,
\begin{equation} \label{lzerolimit}
\lim_{\ell \rightarrow 0}\frac{1}{\sqrt{2\ell+1}}  {\mathbf D}^{\pm} (E, \ell)  = Q^{\pm}(E).
\end{equation}

The quantum Wronskian relation is then rewritten with $\xi$ as,
\begin{align}\label{qWronskian1d}
{\mathbf T}_j(E,\ell)=&  \xi^{-(j+1)}  \mathbf{ D}^{+}(E \Omega^{j+1},\ell) \mathbf{ D}^{-}(E \Omega^{-(j+1)},\ell) \nonumber \\
&-
\xi^{(j+1)} \mathbf{ D}^{-}(E \Omega^{j+1},\ell) \mathbf{ D}^{+}(E \Omega^{-(j+1)},\ell) .
\end{align}

When $q$ is at a root of unity, the SU(2) T-system (\ref{j:Tode}) closes among finite  elements \cite{BLZ1}.
In the present case, this is due to a simple relation,
\begin{equation}\label{truncationk1}
{\mathbf T}_3(E,\ell)=\xi^3 +\xi^{-3} + {\mathbf T}_1(E,\ell).
\end{equation}
Then one ends up with
\begin{align}
{\mathbf T}_1(E \Omega,\ell) {\mathbf T}_1(E\Omega^{-1},\ell) &=1+ {\mathbf T}_2(E,\ell),  \nonumber \\
{\mathbf T}_2(E \Omega,\ell) {\mathbf T}_2(E\Omega^{-1},\ell) &=1+ {\mathbf T}_1(E,\ell) {\mathbf T}_3(E,\ell) \nonumber \\
&=(\xi^3 +{\mathbf T}_1(E,\ell))(\xi^{-3} +{\mathbf T}_1(E,\ell)).
\label{Tsysk1}
\end{align}
In the following, we derive (\ref{N2TBA1}) from the above truncated T-system as the
first nontrivial equation in  the expansion of $h$.
Then  we will show that a similar expansion of  the quantum Wronskian relation  (\ref {qWronskian1})
yields Fendley's solution (\ref{Fendelysolution}).

In this example,  the T-system is identified with the  Y-system.
This is achieved by introducing
\begin{equation}\label{YTk1}
{\mathbf Y}_t(\theta,\ell) ={\mathbf T}_1(E, \ell),   \qquad   {\mathbf Y}_1(\theta, \ell) = {\mathbf T}_2(E,\ell).
\end{equation}
Here  the parameter $\theta$ is related to $E$  by
\begin{equation}\label{Etheta}
E= E_0  {\rm e}^{\frac{2}{3}\theta},
\end{equation}
and the constant $E_0 $ will be determined later.

The Y-system is represented by new variables as
\begin{align}
&{\mathbf Y}_t(\theta-\frac{\pi}{2}i ,\ell)   {\mathbf Y}_t(\theta+\frac{\pi}{2}i,\ell) =1+  {\mathbf Y}_1(\theta,\ell),  \label{Ysystemt}\\
&{\mathbf Y}_1(\theta-\frac{\pi}{2}i ,\ell) {\mathbf Y}_1(\theta+\frac{\pi}{2}i,\ell) =
(\xi^3+  {\mathbf Y}_t(\theta,\ell))  (\xi^{-3}+  {\mathbf Y}_t(\theta,\ell)).  \label{Ysystem1}
\end{align}
Next we consider the expansion in $h$.  
The solution (\ref{trivialT}), strictly at $h=0$,  suggests the expansions 
\begin{equation} \label{yexpansion}
{\mathbf Y}_t(\theta,\ell)= 1+ h y_t(\theta)  +O(h^2),  
\qquad
{\mathbf Y}_1(\theta,\ell)=  h y_1(\theta)   +O(h^2).
\end{equation}
The  first nontrivial equation in the expansion of (\ref{Ysystemt}) is  $O(h)$,
while it is $O(h^2)$  for (\ref{Ysystem1}),
\begin{align}\label{Ysystemexpansionk1}
&y_t(\theta+\frac{\pi}{2}i) + y_t(\theta-\frac{\pi}{2}i) =y_1(\theta),      \nonumber  \\
&y_1(\theta+\frac{\pi}{2}i)y_1(\theta-\frac{\pi}{2}i)=y_t(\theta)^2 +1 .
\end{align}
We set 
\begin{equation}\label{yABrelation}
y_t(\theta)= -B(\theta),  \qquad y_1(\theta)= {\rm e}^{-A(\theta)}
\end{equation} 
and assume that $y_1$ and $y_t$ are  analytic and nonzero in the strip
$\Im m\, \theta \in [-\pi/2,\pi/2]$.
We also assume that the right hand sides of (\ref{yexpansion}) are 
 analytic and nonzero in the narrow strip including the real axis of $\theta$.
These assumptions are  justified by the solution in (\ref{ABsol}), a posteriori.
One then obtains
\begin{align}
A(\theta) &=m_A {\rm e}^{\theta}+C_A -\int_{-\infty}^{\infty} \frac{d\theta'}{2\pi} \frac{1}{\cosh(\theta-\theta') }\ln(1+ B(\theta')^2),  \nonumber \\
B(\theta) &=m_B {\rm e}^{\theta}+C_B -\int_{-\infty}^{\infty} \frac{d\theta'}{2\pi} \frac{1}{\cosh(\theta-\theta')} {\rm e}^{-A(\theta')}  \label{N2TBA2}
\end{align} 
where ``mass terms" are introduced to take account of  the zero mode in the Fourier transformation.
Without loss of generality we can always choose $m_A=1$ by tuning the origin of $\theta$ (or a redefinition of $E_0$).
It will be later shown that $m_B=0$.
The integration constants $C_A$, $C_B$ are found to be zero. 
This can be verified from the asymptotic values 
${\rm e}^{-A(-\infty)}=2/\sqrt{3}, B(-\infty)=-1/\sqrt{3}$.

We have a remark.  The quantum sine Gordon model has ${\cal N}=2$ supersymmetry at a special coupling constant.
Fendley et al. \cite{FI} utilized this and started from the TBA for the generic quantum sine Gordon model.
Then they took a similar limit in the above and derived  (\ref{N2TBA2}).
Here the initial point is different: we start from the ODE.

The above observation concludes that the expansion of ${\mathbf T}_{1,2}$ in $h$  yields  the desired TBA equations (if $m_B=0$).
We then use    (\ref {qWronskian1d}) 
to obtain explicit solutions, given the data
\begin{align}\label{explicitQ}
\lim_{\ell \rightarrow 0} \frac{1}{\sqrt{2\ell+1}} {\mathbf D}^{-} (E, \ell)  &=Q^{-}(E) = \sqrt{\frac{2\pi}{i}}{\rm Ai}(-E),  \nonumber\\
\lim_{\ell \rightarrow 0} \frac{1}{\sqrt{2\ell+1}} {\mathbf D}^{+} (E, \ell)  &=Q^{+}(E)=- \sqrt{\frac{2\pi}{i}}\frac{d}{dE} {\rm Ai}(-E)  .
\end{align}
This looks hopeless  at first sight, as the expansion of the right hand side of (\ref {qWronskian1d})  contains
derivatives of  ${\mathbf D}^{ \pm}$   at $h=0$ which are unknown to us.
By fortunate cancellations of derivative terms, we nevertheless find it possible.
First consider the case $j=0$ in  (\ref {qWronskian1d})   where $\mathbf{T}_0=1$.
The $O(h^0)$ and the $O(h^1)$ equations read respectively,
\begin{align}
&{\rm e}^{-\frac{\pi}{3}i} Q^{+}(E \Omega) Q^{-}(E\Omega^{-1} ) - {\rm e}^{\frac{\pi}{3}i} Q^{-}(E \Omega) Q^{+}(E \Omega^{-1} )=1,  \label{orderh0} \\
&{\rm e}^{\frac{\pi}{3}i}\frac{\partial}{\partial h}
 \bigl( D^{-}(E \Omega,\ell) D^{+}(E \Omega^{-1},\ell )\bigr) |_{h=0}
 - {\rm e}^{-\frac{\pi}{3}i} \frac{\partial}{\partial h}  \bigl( D^{+}(E \Omega,\ell) D^{-}(E \Omega^{-1}, \ell )    )\bigr) |_{h=0}   \nonumber  \\
 &{\phantom{abc}}=
 \frac{i}{3}\Bigl(  {\rm e}^{\frac{\pi}{3}i} Q^{-}(E\Omega) Q^{+}(E\Omega^{-1})+ {\rm e}^{-\frac{\pi}{3}i} Q^{+}(E\Omega) Q^{-}(E\Omega^{-1} )\Bigr).
 \label{orderh1} 
\end{align}
Next consider  $j=1$ in (\ref {qWronskian1d}). 
The $O(h^0)$ term on the right hand side  is found to  be 1 using (\ref{orderh0})
(replacing $E$ by $-E$), 
while the  $O(h^1)$ terms contain derivative terms of $h$.  We find that  these derivative terms can completely  be 
rewritten in terms of $Q^{\pm}$ thanks to (\ref{orderh1}).  Altogether,  one obtains,
\begin{equation}\label{t1expansion}
\mathbf{T}_1(E) = 1-i\Bigl(  {\rm e}^{\frac{\pi}{3}i} Q^{-}(-E\Omega) Q^{+}(-E \Omega^{-1})+ {\rm e}^{-\frac{\pi}{3}i} Q^{+}(-E\Omega) Q^{-}(-E\Omega^{-1} )\Bigr)+O(h^2).
\end{equation}
Thirdly, take $j=2$ in  (\ref {qWronskian1d}).    It is simplified, as $\Omega^3=-1$,
\begin{align}
\mathbf{T}_2(E)&= \xi^{-3} D^+(E \Omega^3,\ell)   D^-(E\Omega^{-3},\ell) -\xi^{3} D^-(E\Omega^3,\ell)   D^+(E\Omega^{-3},\ell)   \nonumber  \\
&= -2ih Q^+(-E) Q^-(-E) +O(h^2).   \label{t2expansion}
\end{align}
Then from equations (\ref{YTk1}),  (\ref{yexpansion}),  (\ref{yABrelation}), (\ref{t1expansion}) and (\ref{t2expansion})
we conclude
\begin{align}\label{ABsol}
B(\theta) &= -i\Bigl(  {\rm e}^{\frac{\pi}{3}i} Q^{-}(-E {\rm e}^{-\frac{\pi}{3}i}) Q^{+}(-E {\rm e}^{\frac{\pi}{3}i})
+ {\rm e}^{-\frac{\pi}{3}i} Q^{+}(-E {\rm e}^{-\frac{\pi}{3}i}) Q^{-}(-E {\rm e}^{\frac{\pi}{3}i})\Bigr) \nonumber \\
&=2\pi \frac{d}{dE}  {\rm Ai}(E {\rm e}^{-\frac{\pi}{3}i})    {\rm Ai}(E {\rm e}^{\frac{\pi}{3}i}),   \nonumber\\
{\rm e}^{-A(\theta)} &= -2i    Q^+(-E) Q^-(-E)  = -2\pi \frac{d}{dE} {\rm Ai}(E)^2
\end{align}
where  we use (\ref{explicitQ}).

Finally, let us check $m_B=0$ in (\ref{N2TBA2})  and evaluate $E_0$ in  (\ref{Etheta}).
This is  done by evaluating the left hand side  of  (\ref{N2TBA2}) in the limit $\theta \rightarrow \infty$ or equivalently, $E \rightarrow \infty$.
The convolution terms do not contribute since  the  integration kernel becomes exponentially small. 
One easily evaluates the asymptotic behavior $\Re e E \gg 1$ from (\ref{ABsol}), 
\[
A(\theta) \sim \frac{4}{3} E^{\frac{3}{2}}  \qquad B(\theta) \sim   - \frac{1}{4} E^{-\frac{3}{2}} +O(E^{-3}).
\]
Using them in the second equation  of (\ref{N2TBA2}),  we conclude $m_B=0$, while the first  equation, with the convention $m_A=1$, leads to
\[
\frac{4}{3} E^{\frac{3}{2}} ={\rm e}^{\theta} \quad  {\rm or} \quad
E=\Bigl(  \frac{3}{4} {\rm e}^{\theta} \Bigr)^{\frac{2}{3}}.
\]
By identifying $z$ with $E$ in  (\ref{Fendelysolution}), we thus conclude that Fendley's solution
is successfully recovered from the ODE.
%
%
\section{Generalizations} \label{sec:generalize}

Let us summarize our findings so far. 
The input is the super-potential $W(x)$.  Once this is fixed,  we  construct a wave function  (\ref{integralrep})
which solves a simple ODE (\ref{AiryODE}). We then ``deform" the ODE by the angular momentum term as in (\ref{AiryWithL}).
This makes the associated Stokes multipliers nontrivial.   Then the first non-trivial response of the 
Stokes multipliers with respect to the small angular momentum  yields Fendley's solution.

Since there exists a list of  super-potentials for ${\cal N}=2$ supersymmetric theories in 2D \cite{VafaWarner, Martinec, CV},
one naturally wonders if the above procedure works,  starting  from other super-potentials.
Below we will discuss the super-potential of the type SU(2)$_k$  and SU(3)$_1$
which provide the affirmative evidences to this expectation.
%
%
\subsection{Exact solution : SU(2)$_k$ }
In this case, the relevant super-potential takes the form,
\[
W_k(x={\rm e}^{i\theta} + {\rm e}^{-i\theta} )= \frac{2}{k+2} \cos (k+2) \theta.
\]
More explicitly,
\begin{align*}
W_1(x)&= \frac{x^3}{3}-x,&
W_2(x) &=  \frac{1}{4}(x^4-4 x^2+2),  \\
W_3(x) &= \frac{1}{5}(x^5-5 x^3+5 x),&
W_4(x)&=\frac{1}{6}(x^6-6 x^4+9 x^2 -2)
\end{align*}
and so on.\pn
According to the above strategy we first construct a wave function,
\begin{equation}
\phi^{(k)}(x, E)= \int_{\cal C} {\rm e}^{ (x-E)^{\frac{k+2}{2}} W_k(z (x-E)^{-1/2}) } dz.
\label{integralrepphik}
\end{equation}
The contour should meet the requirement that $\phi^{(k)}(x,E)$ is exponentially decreasing on the real axis of $x$.

Note that  (\ref{integralrep}) is contained as the $k=1$ case.
We immediately see that  $\phi^{(k)}(x,E)$ satisfies an ODE,
\begin{equation}\label{su2kODE}
\Bigl( -\frac{d^2}{dx^2} +(x-E)^k \Bigr) \phi^{(k)}(x,E) =0.
\end{equation}

As before, we interpret this as  the  $(n,\alpha)=(2,\frac{1}{2})$ case  of
a  generalized ODE, 
\begin{equation}\label{su2kODEII}
\Bigl( -\frac{d^2}{dx^2} +(x^{2\alpha}-E)^k \Bigr) \phi^{(k)}(x, E) =0,
\end{equation} 
with  vanishing  boundary condition for $\Re e\, x \gg 1$. \pn 
This has been proposed to be the ODE for the spin $\frac{k}{2}$ SU(2) case 
with $q={\rm e}^{i\pi/(\alpha k+1)}$ \cite{Lukyanovup, DDMST}.

We have, analogously to (\ref{explicitQpmk1}),
\begin{align}\label{explicitQpmk}
Q^-(E)  &=   \sqrt{  \frac{2E i}{(k+2)\pi }  }   K_{\frac{1}{k+2}} \bigl(   \frac{2}{k+2}(-E)^{\frac{k+2}{2}} \bigr), \\
Q^+(E)  &=  -  \frac{d}{dE}  Q^-(E) 
\end{align}
where $K_{\nu}$ stands for the modified Bessel function.
We then include the angular momentum term with the effect, 
\[
Q^-(E) \rightarrow  D^-(E,\ell),  \qquad 
Q^+(E) \rightarrow  D^+(E,\ell) .
\]

The Stokes multiplier takes the same form  as (\ref{DVF3}), while the parameters take different values,
\begin{equation}\label{OmegaXik}
\Omega={\rm e}^{-\frac{\pi}{k+2}i},   \qquad \xi={\rm e}^{\frac{\pi-h}{k+2} i}.
\end{equation}

The SU(2) T system remains valid, while 
(\ref{truncationk1}) is  replaced by
\[
{ \mathbf  T}_{k+2}(E,\ell)=\xi^{k+2}+ \xi^{-(k+2)}+{ \mathbf  T}_k(E,\ell).
\]
The transformation from the T-system to the Y-system is accomplished by \cite{KluemperPearce},
\begin{align}
{ \mathbf Y}_j(\theta,\ell)&={ \mathbf T}_{j-1}(E,\ell){ \mathbf T}_{j+1}(E,\ell)   \qquad (1\le j\le k),  \nonumber \\
\qquad
{ \mathbf Y}_t(\theta,\ell)&={ \mathbf T}_{k}(E,\ell),    \label{KPtransformation}  \\
&E=E_0^{(k)}\, {\rm e}^{\frac{2}{k+2} \theta}  \label{EThetak}
\end{align}
where ${ \mathbf T}_0$ is set to be 1.
  The coefficient $E_0^{(k)}$ will be determined later.

We obtain as a result,
\begin{align}\label{Ysystemk}
{ \mathbf Y}_j(\theta+\frac{\pi}{2}i ,\ell){ \mathbf Y}_j(\theta+\frac{\pi}{2}i ,\ell)
&=(1+ { \mathbf Y}_{j-1}(\theta,\ell))(1+ { \mathbf Y}_{j+1}(\theta,\ell))   \quad (1\le j \le k-1), \nonumber   \\
{ \mathbf Y}_k(\theta+\frac{\pi}{2}i ,\ell){ \mathbf Y}_k(\theta+\frac{\pi}{2}i ,\ell)
&=(1+ { \mathbf Y}_{k-1}(\theta,\ell))(1+\xi^{k+2} { \mathbf Y}_{t}(\theta,\ell))(1+\xi^{-(k+2)} { \mathbf Y}_{t}(\theta,\ell)),   \nonumber \\
{ \mathbf Y}_t(\theta+\frac{\pi}{2}i ,\ell){ \mathbf Y}_t(\theta+\frac{\pi}{2}i ,\ell)
&=(1+ { \mathbf Y}_{k}(\theta,\ell))  
\end{align}
where we set ${ \mathbf Y}_0=0$.

By strictly setting $h=-2\ell\pi=0$,  we obtain constant solutions for $t_{0,j}={\mathbf T}_j(E,0)$ 
\[
t_{0,j}= \frac{\sin\frac{(j+1)\pi}{k+2}}{\sin\frac{\pi}{k+2}} \qquad 1\le j \le k+1 
\]
especially $t_{0,k+1}=0$.
Then ${\mathbf Y}_j(\theta,0)$ is  determined by (\ref{KPtransformation}).
We  assume the expansion around $h=0$,
\begin{align}\label{TYhexpansion}
{\mathbf T}_j(\theta,\ell) &= t_{0,j} + h t_{1,j}(\theta) +O(h^2)    \qquad (1\le j \le k+1 ),  \nonumber \\
{\mathbf Y}_j(\theta,\ell) &=y_{0,j}+ h y_{1,j}(\theta)  +O(h^2)   \qquad  (1\le j \le k),   \nonumber \\
{\mathbf Y}_t(\theta,\ell) &= 1+h y_{1,t}(\theta)  +O(h^2)  .
\end{align}
Note that  $y_{0,j} \ne 0$ if $j<k$ and
${\mathbf Y}_k(\theta,\ell)= h y_{k}(\theta)$.

This leads to an important consequence.
Although the  original Y system (\ref{Ysystemk}) consists of  $k+1$ equations among $k+1$ Y functions,
the  first nontrivial relations {\it close} only  among $y_{1,k}(\theta)$ and $y_{1,t}(\theta)$,
\begin{align*}
y_{1,k} (\theta+\frac{\pi}{2}i) y_{1,k} (\theta-\frac{\pi}{2}i)  &=( t_{0,k-1})^2  (y_{1,t}(\theta)^2+1),  \\
y_{1,t}(\theta+\frac{\pi}{2}i) + y_{1,t}(\theta-\frac{\pi}{2}i)  &= y_{1,k} (\theta).
\end{align*}

They are very similar to (\ref{Ysystemexpansionk1}) by identifying $y_{1,k}=y_1$.
Consequently,  one obtains the analogous TBA 
\begin{align}
A(\theta) &=m_A {\rm e}^{\theta}- \ln 2 \cos \frac{\pi}{k+2}  
  -\int \frac{d\theta'}{2\pi} \frac{1}{\cosh(\theta-\theta') }\ln(1+ B(\theta')^2),  \nonumber \\
B(\theta) &=m_B {\rm e}^{\theta} -\int \frac{d\theta'}{2\pi} \frac{1}{\cosh(\theta-\theta')} {\rm e}^{-A(\theta')}  \label{N2TBAk}
\end{align} 
where 
\begin{equation}\label{yABrelationk}
y_{1,t}(\theta)= -B(\theta),  \qquad y_{1,k}(\theta)= {\rm e}^{-A(\theta)}.
\end{equation} 
The integration constants are fixed by the asymptotic values,
\[
{\rm e}^{-A(-\infty)} = 2 \cot \frac{\pi}{k+2}, 
\qquad 
B(-\infty) = -\cot \frac{\pi}{k+2}.
\]
As before we choose $E^{(k)}_0$ in (\ref{EThetak})  such that  $m_A=1$.
Below we will argue that $m_B=0$.  Then the resultant TBA agrees with the result in \cite{CFIV}
(for $\Theta={\pi}/{k+2}$).
We again remark that (\ref{N2TBAk}) was derived in \cite{CFIV}  by taking a limit 
from the TBA for the ${\cal N}=0$ sine Gordon model at a special coupling constant 
which consists of $k+1$ integral equations.

Now we are in position to 
derive the explicit solutions for $A(\theta)$ and $B(\theta)$ or equivalently, $y_{1,k}(\theta)$ and  $y_{1,t}(\theta)$.
Thanks to (\ref{KPtransformation}) and (\ref{TYhexpansion}), one immediately finds
\begin{equation}
y_{1,t}(\theta)=t_{1,k}(E),  \qquad
y_{1,k}(\theta)=t_{0,k-1} t_{1,k+1}(E).
\end{equation}

The right hand side of above equations can be evaluated through the quantum Wronskian relations
(\ref{qWronskian1d}) with $j=0, k$ and $k+1$.
Note that $\Omega$ and $\xi$ are given in (\ref{OmegaXik}).
After simple manipulations, we obtain,
\begin{align*}
y_{1,k}(\theta) &=-2 i t_{0,k-1}  Q^-(-E) Q^+(-E), \\
y_{1,t}(\theta)&=-i\Bigl( {\rm e}^{\frac{\pi}{k+2}i} Q^-(-E \Omega) Q^+(-E \Omega^{-1}) + 
{\rm e}^{-\frac{\pi}{k+2}i} Q^-(-E \Omega^{-1}) Q^+(-E \Omega)\Bigr).
\end{align*}

By the use of (\ref{explicitQpmk}),  solutions are represented  explicitly.
Let
\begin{equation}
{\rm Ai}^{(k)}(E)= \frac{1}{\pi} \sqrt{\frac{E}{k+2}} K_{\frac{1}{k+2}}(\frac{2}{k+2} E^{\frac{k+2}{2}})
\end{equation}
which reduces to the Airy function ${\rm Ai}$ if $k=1$.
Then   one finds solutions which generalize  (\ref{ABsol}) naturally, 
\begin{align}\label{ABsolk}
B(\theta) 
&=2\pi \frac{d}{dE}  {\rm Ai}^{(k)} (E \Omega )    {\rm Ai}^{(k)}(E  \Omega^{-1}),   \nonumber\\
{\rm e}^{-A(\theta)} &=  -4\pi \cos \frac{\pi}{k+2}  \frac{d}{dE} {\rm Ai}^{(k)}(E)^2.
\end{align}

Then we substitute  the above explicit solution into (\ref{N2TBAk}) and take the limit   $\theta \rightarrow \infty$.
From the known asymptotic behavior
\[
{\rm Ai}^{(k)}(E)  \sim \frac{E^{-\frac{k}{4}}}{2 \sqrt{\pi}}
 {\rm e}^{-\frac{2}{k+2} E^{\frac{k+2}{2}}},
\]
one easily  checks that $m_B=0, m_A=1$ and
\[
E= E_0^{(k)} {\rm e}^{\frac{2}{k+2}\theta},  \qquad  E_0^{(k)}=\Bigl(\frac{k+2}{4}\Bigr)^{\frac{2}{k+2}}.
\]
We assume  $k \in \mathbb{N}$ in deriving  TBA  (\ref{N2TBAk}).
Once  it is obtained,  however, $k$ enters as a  mere parameter.  
One can take, for example, $k \in \mathbb{R}_{\ge 0}$. 
We checked numerically that  (\ref{ABsolk})  still satisfies (\ref{N2TBAk}) in this case.
%
%
\subsection{Exact solution : SU(3)$_1$ }
Next consider the perturbation of  the  SU(3)$_1$ type.
The super-potential  reads,
\[
W(z,x)= \frac{z^4}{4}- x z.
\]
According to our working hypothesis, we introduce  
\[
\phi(x, E)= \int_{{\cal C}} {\rm e}^{W(z,x-E)} dz
\]
which satisfies the $3^{\text rd}$ order ODE,
\begin{equation} \label{ODEsu30}
\Bigl( \frac{d^3}{dx^3} + x-E\Bigr) \phi(x,E) =0.
\qquad 
\end{equation}
As always,  we choose ${\cal C}$ such that  the asymptotic behavior of $\phi$ agrees with (\ref{j:asym}).

We regards this as a special case of 
(\ref{j:Schr}),
with  $n=3$ and $\alpha=1/3$.  
The analysis in \cite{DTsu3, BHK}  shows that the ODE  is related to CFT with $A^{(2)}_2$ symmetry.

%

The explicit solution to (\ref{ODEsu30}) with the desired property (\ref{j:asym}) 
is given by Meijer's $G$ function or a linear combination of 
confluent hypergeometric series as discussed in \ref{appendixodesolution}~~.
 Here we do not specify its explicit form but use a symbol $\varphi$: 
\begin{equation}\label{defvarphi}
\varphi(x-E)=\phi(x,E).
\end{equation}

We define, for later use, 
\begin{align} \label{Qsu3}
Q^{[0]}(E)&=\varphi(-E),&
Q^{[1]}(E)&=-\frac{d}{dE} \varphi(-E),  \nonumber \\
Q^{[2]}(E)&=\frac{1}{2}\frac{d^2}{dE^2} \varphi(-E)
\end{align}

As before, we  consider  a ``radial" ODE  \cite{DDTsuN}
\begin{equation}\label{ODEsu3}
\Bigl(
{\cal D}(g_2-2) {\cal D}(g_1-1){\cal D}(g_0) +x-E
\Bigr) \psi(x,E, {\bf g})=0.
\end{equation}
The operator  ${\cal D}(g)$ is defined by
 \[
{\cal D}(g) :=  \frac{d}{dx} -\frac{g}{x}.
\]
The parameters $g_i$ are constrained by
\[
g_0+g_1+g_2=3.
\]
We denote by $\phi(x,E,{\bf g})$,
 the solution which behaves as
(\ref{j:asym})   in ${\cal S}_0$ where ${\bf g}$ stands for $\{g_0,g_1,g_2\}$.
We further introduce  
\begin{equation}\label{yjsu3}
 \phi_j(x, E, {\bf g})= q^j \phi(q^{-j} x ,  \Omega^{3j} E,{\bf g})
\end{equation}
where $q={\rm e}^{2\pi i/(3\alpha+3)}$ and   $\Omega=q^{-\alpha}$
\footnote{We are now considering $\alpha=\frac{1}{3}$
thus explicitly $q=\Omega^{-3}={\rm e}^{i\pi/2}$}.
Then any $ \phi_j(x, E,{\bf g}), j \in \mathbb{Z}$ is also a solution to the ODE.

The connection relations among $\{\phi_j\}$ remain formally the  same  as (\ref{ODEconnection}),
although the  components in (\ref{Stokesmatrix}) now possess a dependency on ${\bf g}$.

We denote by $\{\chi^{[i]}\}$, another FSS of (\ref{ODEsu3}) near the origin characterized by the behavior for $ x\rightarrow 0$,
\[
\chi^{[i]}(x, E, {\bf g} ) \sim  {\cal N}_g  x^{g_i}   \, \, (i=0,1,2), 
\qquad   {\cal N}_g = \bigl(\prod_{0\le j<i \le 2} (g_i-g_j)\bigr)^{-\frac{1}{3}}.
\]
The ordering, $\Re{\rm e} \,g_0 < \Re{\rm e} \,g_1<\Re{\rm e} \,g_2$  is assumed from now on.
The normalization factor  ${\cal N}$ is chosen so that
the Wronskian determinant of $\{\chi^{[0]},\chi^{[1]},\chi^{[2]} \}$ is unity.
Following (\ref{yjsu3}), we set
\[
\chi^{[i]}_j  (x,E,{\bf g})= q^j \chi^{[i]}(q^{-j} x, \Omega^{3j}E,{\bf g})= q^j \chi^{[i]}(q^{-j} x, q^{-j}E,{\bf g}) .
\]
It can be shown as in the case of SU(2) \cite{DT99}, 
\[
\chi^{[i]}_j  (x,E,{\bf g})=q^{j(1-g_i)}  \chi^{[i]}(x,E,{\bf g}).
\]

We set analogously to  (\ref{radialconnection}), 
\[
\phi(x, E, {\bf g})= \sum_{i=0}^2  D^{[i]}(E,\mathbf{g})  \chi^{[i]}(x, E, {\bf g} )
\]
or slightly more generally,
\begin{equation}\label{radialsu3}
 \phi_j (x, E, {\bf g})= \sum_{i=0}^2  D^{[i]}(E q^{-j} ,\mathbf{g})  \chi^{[i]}_j (x, E, {\bf g} ).
\end{equation}
When $\{g_0,g_1,g_2\}=\{0,1,2\}$ the original ODE  (\ref{ODEsu30}) is recovered.
This  implies the limit
\begin{equation}\label{limitsu3}
\lim_{
\{g_0,g_1,g_2\}\rightarrow\{0,1,2\}}    {\cal N}_g  D^{[i]}(E,\mathbf{g})   =Q^{[i]}(E).
\end{equation}

The generalized Stokes multipliers
\begin{equation}\label{wronskiansu3}
\tau^{(1)}_m(E,{\bf g})=W^{(3)}_{0,m+1,m+2}(E,{\bf g}),  \qquad
\tau^{(2)}_m(E,{\bf g})=W^{(3)}_{0,1,m+2} (E,{\bf g})
\end{equation}
are  expressible in terms of  $ D^{[i]}(E,\mathbf{g})$, using (\ref{radialsu3}) in the above,

\begin{align}\label{qWronskiansu3}
\tau^{(1)}_m(E,{\bf g}) =& 
 \sum_{\sigma} {\rm sgn}\sigma q^{(m+1)(1-g_{\sigma_2})}q^{(m+2)(1-g_{\sigma_3})}   \nonumber \\
 &\times 
  D^{[\sigma_1]} (E,{\bf g})  D^{[\sigma_2]} (E q^{-(m+1)},{\bf g}) D^{[\sigma_2]} (E q^{-(m+2)},{\bf g}),  \nonumber \\
  \nonumber  \\
\tau^{(2)}_m(E,{\bf g}) =&
 \sum_{\sigma} {\rm sgn}\sigma q^{(1-g_{\sigma_2})}q^{(m+2)(1-g_{\sigma_3})}   \nonumber \\
&\times   D^{[\sigma_1]} (E,{\bf g})  D^{[\sigma_2]} (E q^{-1},{\bf g}) D^{[\sigma_2]} (E q^{-(m+2)},{\bf g}) 
\end{align}
where $\sigma$ signifies the permutation of $\{0,1,2\}$.

They also have the  DVF representations \cite{DDTsuN}.
The explicit forms are given  in \ref{appendixDVFsu3} ~ for $\tau^{(1)}_1$ and  $\tau^{(2)}_1$.
The results there suggest that it is convenient to  set,
\begin{equation}\label{Ttaurelation}
{\bf  T}^{(a)}_m(E)= \tau^{(a)}_m(E q^{\frac{a+m}{2} +\frac{1}{4}},{\bf g})
\end{equation}
and 
\begin{equation}\label{EthetasuN}
E = E_0 {\rm e}^{\frac{3}{4}\theta}.
\end{equation}
 The ${\bf g}$ dependency is dropped in $\mathbf{T}$.
The SU(3) T-system is then  recovered,
\begin{equation}
{\mathbf T}^{(a)}_m(\theta+\frac{\pi}{3}i ){\mathbf T}^{(a)}_m(\theta-\frac{\pi}{3}i )=
{\mathbf T}^{(a+1)}_m(\theta) {\mathbf T}^{(a-1)}_m(\theta) +{\mathbf T}^{(a)}_{m+1}(\theta)  {\mathbf T}^{(a)}_{m-1}(\theta)
\quad 
\label{Hirotasu3}
\end{equation}
where   $ a=1,2$ and $ m \ge 1$.
We set   ${\bf  T}^{(0)}_m ={\bf  T}^{(3)}_m=1$ and ${\bf  T}^{(a)}_0=1$
and  used $q={\rm e}^{\frac{\pi}{2} i}$.
We perform a transformation  similar to  (\ref{YTtransformation})  \cite{KNS},
\begin{equation}\label{YTsu3}
{\bf  Y}^{(a)}_m(\theta) =
\frac{   {\bf  T}^{(a)}_{m-1}(\theta) {\bf  T}^{(a)}_{m+1}(\theta)}
        {   {\bf  T}^{(a-1)}_{m}(\theta) {\bf  T}^{(a+1)}_{m}(\theta)}.
\end{equation}
This yields the SU(3) Y system
\begin{equation}
{\bf  Y}^{(a)}_m(\theta+\frac{\pi}{3}i ) {\bf  Y}^{(a)}_m(\theta+\frac{\pi}{3}i )=
\frac{ (1+{\bf  Y}^{(a)}_{m-1}(\theta))(1+{\bf  Y}^{(a)}_{m+1}(\theta))}
{   (1+({\bf  Y}^{(a+1)}_{m}(\theta))^{-1})(1+({\bf  Y}^{(a-1)}_{m}(\theta))^{-1})     }   
\label{Ysystemsu3}
\end{equation}
where   $ a=1,2$, $ m \ge 1$
and set  $({\bf  Y}^{(0)}_m)^{-1} =({\bf  Y}^{(3)}_m)^{-1}=0$,
${\bf  Y}^{(a)}_0=0$.
Note that, contrary to the SU(2) case, the truncation of  the Y system 
(\ref{Ysystemsu3}) to a finite set does not occur
\footnote{
There is, however, an elaborate way to introduce a set of
nonlinear equations which truncate among finite elements \cite{DamerauKluemper}.
We however do not adopt that approach here.}.
From now on, we consider the special choice on $g_i$,
\begin{equation}\label{gchoice}
g_0=\frac{h}{2\pi} , \quad
g_1=1+\frac{h}{2\pi}   , \quad
g_2=2-\frac{h}{\pi}
\end{equation}
and take the limit $h \rightarrow 0$.

We can easily check (for $a=1,2$,  $k \in \mathbb{Z}_{\ge 0}$ )
\begin{equation}
\lim_{h \rightarrow 0}  {\bf  T}^{(a)}_{4k}=\lim_{h \rightarrow 0}  {\bf  T}^{(a)}_{4k+1}=1, \qquad 
\lim_{
h \rightarrow 0}  {\bf  T}^{(a)}_{4k+2}=\lim_{
h \rightarrow 0}  {\bf  T}^{(a)}_{4k+3}=0.
\end{equation}

This motivates us to assume the expansions, 
\begin{align}
 {\bf  T}^{(a)}_{m} (\theta)&=1  + h t^{(a)}_{1,m}(\theta) +h^2  t^{(a)}_{2,m}(\theta)+ O(h^3) \qquad
 m=4k,4k+1,    \nonumber \\
 {\bf T}^{(a)}_{m} (\theta)&=  h t^{(a)}_{1, m} (\theta)+ h^2  t^{(a)}_{2,m}(\theta)+ O(h^3)
 \qquad
 m=4k+2,4k+3.   \label{Texpansionsu3}
\end{align}
Similarly consider the expansions of the Y function,
\begin{equation}
{\bf  Y}^{(a)}_m(\theta)= y^{(a)}_{0,m} + h y^{(a)}_{1,m}(\theta)  + h^2 y^{(a)}_{2,m}(\theta) +O(h^3).
\end{equation}

By using the $k=1$ case of (\ref{Texpansionsu3})  in  (\ref{YTsu3})  we obtain
\begin{equation}\label{y1expansion}
{\bf  Y}^{(a)}_1(\theta)= h t^{(a)}_{1,2} (\theta)+
h^2 ( -t^{(a)}_{1,2} (\theta)  t^{(\bar{a})}_{1,1} (\theta)+ t^{(a)}_{2,2} (\theta) )+O(h^3) 
\end{equation}
for $(a,\bar{a})=(1,2)$ or $(2,1)$.
Substituting these into  (\ref{Ysystemsu3}),  one deduces expansions for
 other  Y functions,
\begin{align}\label{otheryexpansions}
{\bf  Y}^{(a)}_2(\theta) &=-1+ h y^{(a)}_{1,2}(\theta)   +  h^2 y^{(a)}_{2,2} (\theta)+  O(h^3),    \nonumber  \\
{\bf  Y}^{(a)}_3(\theta) &=-1+  h y^{(a)}_{1,3}(\theta)   +  h^2 y^{(a)}_{2,3}(\theta) +  O(h^3)      \qquad (a=1,2)  
\end{align}
and so on.
There are complex expressions of 
$y^{(a)}_{j,m}$  in terms of
$t^{(a')}_{j',m'}$, which we shall omit.

The first nontrivial relations of the case  $m=1$ in (\ref{Ysystemsu3})
exist at  O($h^2$),

\begin{equation}\label{epsiloneq1}
y^{(a)}_{1,1} ( \theta+\frac{\pi}{3}i)
 y^{(a)}_{1,1} ( \theta-\frac{\pi}{3}i)
 = y^{(\bar{a})}_{1,1} (\theta)   y^{(a)}_{1,2} (\theta) 
\end{equation}
where $(a,\bar{a})=(1,2)$  or  $(2,1)$.

Next consider the $m=2$ case.
The O($h^0$) equations  require
\begin{equation}\label{etarestriction}
y^{(a)}_{1,3} (\theta)=- y^{(\bar{a})}_{1,2}(\theta)
\end{equation}
 while  the  O($h^1$) equations yield
\begin{equation}\label{etaeq1}
y^{(a)}_{1,2} (\theta+\frac{\pi}{3}i) + y^{(a)}_{1,2}  (\theta-\frac{\pi}{3}i) 
= - y^{(a)}_{1,1} (\theta)  +y^{(\bar{a})}_{1,2}(\theta)+
\frac{  y^{(a)}_{2,3}(\theta) + y^{(\bar{a})}_{2,2}(\theta)}{y^{(\bar{a})}_{1,2}(\theta)}.
\end{equation}


This again significantly differs from the SU(2) case.
The equations do not  close among  $y^{(a)}_{1,m}$, i.e., the first order coefficients in $h$.
To determine the last terms in  (\ref{etaeq1}), one must consider equations containing  $y^{(a)}_{3,m}$
and so on.  This leads to an  infinite hierarchy of equations.

There are, however,  miraculous cancellations.   In  \ref{appendixCancellation} ,
it will be  shown  that  as a direct consequence of (\ref{qWronskiansu3}),
 the last terms in  (\ref{etaeq1}) are simplified drastically,
 \begin{equation}\label{simplification}
 \frac{  y^{(a)}_{2,3}(\theta) + y^{(\bar{a})}_{2,2}(\theta)}{y^{(\bar{a})}_{1,2}(\theta)}
 =  \begin{cases}
  3i&   a=1, \\
 -3i&  a=2 .
 \end{cases}
\end{equation}

Thus  equations (\ref{epsiloneq1}) and (\ref{etaeq1})  provide the closed relations among
$y^{(a)}_{1,m}\, (a,m =1,2)$.

The result can be neatly written down by introducing, 
\begin{equation}\label{ABdef}
y^{(a)}_{1,1}(\theta)  ={\rm e}^{- A_a(\theta)}, 
 \quad
 y^{(1)}_{1,2}(\theta)  = B_0(\theta) + i,\quad
 y^{(2)}_{1,2}(\theta)  =B_{\bar{0}}(\theta) - i.
\end{equation}

Then we have 
\begin{align*}
&{\rm e}^{-A_1(\theta+\frac{\pi}{3}i) -A_1(\theta-\frac{\pi}{3}i) + A_{2}(\theta)}
= B_0(\theta) + i, \\
&{\rm e}^{-A_2(\theta+\frac{\pi}{3}i) -A_2(\theta-\frac{\pi}{3}i) + A_{1}(\theta)}
= B_{\bar{0}}(\theta) - i, \\
&B_0(\theta+\frac{\pi}{3}i)+ B_0(\theta-\frac{\pi}{3}i) -B_{\bar{0}}(\theta) =-{\rm e}^{-A_1(\theta)}, \\
&B_{\bar{0}}(\theta+\frac{\pi}{3}i)+ B_{\bar{0}}(\theta-\frac{\pi}{3}i) -B_{0}(\theta) =-{\rm e}^{-A_2(\theta)}.\\
\end{align*}

Under suitable assumptions on analyticity, we obtain the following TBA 
\begin{align}\label{TBAsu3}
%
%
A_r(\theta)&=m_r {\rm e}^{\theta}  -\sum_{\ell =0,\bar{0}}
\int_{-\infty}^{\infty}   \frac{d\theta'}{2\pi}\Phi_{r,\ell} (\theta-\theta') \ln (i a_{\ell}+B_{\ell}(\theta') ) \quad(r=1,2), \\
B_{\ell}(\theta)&=  -\sum_{r=1,2 }
\int_{-\infty}^{\infty}   \frac{d\theta'}{2\pi}\Phi_{r,\ell} (\theta-\theta'){\rm e}^{ -A_r(\theta') }  \quad(\ell=0,\bar{0}) \nonumber 
\end{align}
where  $a_0=-a_{{\bar 0}}=1$ and 
\begin{align*}
\Phi_{1,0}(\theta)=& \Phi_{2,\bar{0}}(\theta)=\frac{\sin\frac{\pi}{3} }{\cosh \theta -\cos\frac{\pi}{3}},  \\
\Phi_{1,\bar{0}}(\theta)=& \Phi_{2,0}(\theta)=\frac{\sin\frac{\pi}{3} }{\cosh \theta +\cos\frac{\pi}{3}}.
\end{align*}
We have used  the limiting values
\begin{align*}
{\rm e}^{-A_1(-\infty)} &= \frac{3}{2\sqrt{2}}{\rm e}^{\frac{\pi}{4}i}, &
{\rm e}^{-A_2(-\infty)} &= \frac{3}{2\sqrt{2}} {\rm e}^{-\frac{\pi}{4}i},  \\
B_0(-\infty)&=-\frac{3+i}{4}, &
B_{\bar{0}}(-\infty)&=-\frac{3-i}{4} 
\end{align*}
to fix the integration constants.   The mass coefficients $m_r$ can be set to unity with proper choice of $E_0$ in 
(\ref{EthetasuN}).

Next let us discuss the solutions in terms of $\varphi$  in (\ref{defvarphi}).  
Our strategy is similar to the SU(2) case.
Expand the Wronskian relations (\ref{wronskiansu3}) in powers of $h$.
Use the fact $\tau^{(1)}_0=1$ to replace the derivatives of
 $D^{[i]}$ by  $Q^{[i]}$  taking (\ref{limitsu3}) into account.   
 Then use  (\ref{Qsu3})
 and represent the result by $\varphi$.  
 There is, of course, no guarantee that all derivatives can be rewritten by this trick.
We however found that, parallel to the SU(2) case, this replacement can be
done successfully.

As the analogous argument is  presented for (\ref{simplification}) in  \ref{appendixCancellation} , we shall 
omit  details and write down the final  results,
\begin{align}\label{su3solution}
{\rm e}^{-A_1(\theta)} &=t^{(1)}_{1,2}(\theta)  
%
=3 \omega^3  w_E[\varphi(E\omega^{-1}) , \varphi(E\omega^{3})] \frac{d^2}{dE^2} \varphi(E \omega^{-1}), 
\nonumber\\
{\rm e}^{-A_2(\theta)}& =t^{(2)}_{1,2}(\theta)  
%
=3 \omega^{-3}  w_E[\varphi(E\omega^{-3}) , \varphi(E\omega)] \frac{d^2}{dE^2} \varphi(E \omega)
\end{align}
where $\omega={\rm e}^{\frac{\pi}{8}i}$ and 
$w_E[f,g]= f \frac{d}{dE} g- g \frac{d}{dE} f$.

We use (\ref {solB0B0b})  in  \ref{appendixCancellation}~~~to write down the solutions for $B_0$ and $B_{\bar{0}}$,
\begin{align}\label{solB0B0bmain}
B_{\bar{0}} (\theta)&=
y^{(2)}_{1,2}(\theta)+i     
%
=3 \omega^{-1} w_E[\varphi(E\omega^{-1}) , \varphi(E\omega^{-5})]  \frac{d^2}{dE^2} \varphi(E \omega^3) +i, 
\nonumber \\
B_{0} (\theta)&=
y^{(1)}_{1,2}(\theta) -i    
%
=-3\omega   w_E[\varphi(E\omega) , \varphi(E\omega^{5})]  \frac{d^2}{dE^2} \varphi(E \omega^{-3}) -i.
\end{align}

By using the $n=3$ case of  (\ref{j:asym})  
 in  (\ref{su3solution}) and (\ref {solB0B0bmain}),
 the $\theta \rightarrow \infty$ asymptotic forms of 
 $A_r, B_{0}$ and $B_{\bar{0}}$  are easily derived.   
 Substituting these into   (\ref{TBAsu3}), we can fix $E_0$ in   (\ref{EthetasuN}) (with $m_r=1$)
  \[
 E_0= \Bigl(  \frac{4}{3\sqrt{3}}\Bigr)^{\frac{3}{4}} .
  \]

%
\section{Applications}  \label{sec:application}
\subsection{Analytic evaluation of the  Cecotti-Fendley-Intriligator-Vafa   Index}
We have been able to derive some explicit solutions for ${\cal N}=$2 TBA.
This may open up the possibility to investigate these systems in a quantitative manner.
As an application of the above results, we discuss the analytic evaluation of 
the Cecotti-Fendley-Intriligator-Vafa  (CFIV) index  ${\mathbf Q}_{\rm CFIV}$ \cite{CFIV} for the SU(2)$_k$ case
in the massless limit.
The CFIV index is defined by
\[
{\rm Tr} (-1)^F F {\rm e}^{-\beta H}
\]
where $F$ denotes the Fermion number.  
This is reformulated in the TBA framework and it is expressed as
\[
{\mathbf Q}_{\rm CFIV}(\mu)= \mu \int_{-\infty}^{\infty}  \frac{d\theta}{2\pi} \cosh \theta {\rm e}^{-A(\theta,\mu)}
\]
where $\mu=m \beta$ and $m$ is the mass. 
In the view point of $tt^{\ast}$ geometry, $\mu$ corresponds to the radial coordinate 
and  ${\mathbf Q}_{\rm CFIV}(\mu)$ is expressible in terms of the solution to the PIII equation.
We are interested in  the massless limit.
In this case ${\mathbf Q}_{\rm CFIV}$ is a constant and given by
\[
{\mathbf Q}_{\rm CFIV}=  2\int_{-\infty}^{\infty} \frac{d\theta}{2\pi}{ \rm e}^{ \theta} {\rm e}^{-A(\theta)}
\]
where the factor 2 takes account of the contributions from the left and right edges.
By change of integration variables from $\theta$ to $E$, we have
\begin{align*}
{\mathbf Q}_{\rm CFIV}&= 2\int_{0}^{\infty} dE E^{\frac{k}{2}} {\rm e}^{-A(\theta)} \\
&=16 \pi \cos \frac{\pi}{k+2}  \int_0^{\infty} dE   E^{\frac{k}{2}}  {\rm Ai}^{(k)}(E)   \frac{d}{dE}  {\rm Ai}^{(k)}(E).
\end{align*}
Using the recursion relation for the modified Bessel function,  $\frac{d}{dE} {\rm Ai}^{(k)}(E) $ 
is given by a sum of three terms.
We introduce the notation,
\[
{\cal G}_n(p,q,\ell )
=\int_0^{\infty} dE E^{2n+\ell+1} K_{\frac{p}{n+1}} (\frac{E^{n+1}}{n+1})K_{\frac{q}{n+1}} (\frac{E^{n+1}}{n+1}).
\]
Then ${\mathbf Q}_{\rm CFIV}$ is expressed as
\[
{\mathbf Q}_{\rm CFIV}= \frac{8}{\pi(k+2)}({\cal G}_{\frac{k}{2}} (\frac{1}{2},\frac{1}{2},\-\frac{k}{2}-1 )
-{\cal G}_{\frac{k}{2}}(\frac{1}{2},-\frac{k+1}{2},0)-{\cal G}_{\frac{k}{2}}(\frac{1}{2}, \frac{k+3}{2},0)).
\]
On the other hand, ${\cal G}_n(p,q,\ell )$ is already evaluated in \cite{Cecottietal},
\begin{align}\label{Cecottiformula}
{\cal G}_n(p,q,\ell ) =&   
\frac{ (2(n+1))^{1+\frac{\ell}{n+1}}  }{   4 \Gamma(2+\frac{\ell}{n+1})  }
\Gamma( 1+\frac{p+q+\ell}{2(n+1)})\Gamma( 1+\frac{-p+q+\ell}{2(n+1)})  \nonumber \\
&\times \Gamma( 1+\frac{p-q+\ell}{2(n+1)})
\Gamma( 1+\frac{-p-q+\ell}{2(n+1)}).
\end{align}
Substituting these results, we find  that the resultant  ${\mathbf Q}_{\rm CFIV}$ is simplified considerably,
\[
{\mathbf Q}_{\rm CFIV}=  \frac{k}{k+2}
\]
which agrees with the known result in \cite{CFIV}.
\subsection{Sub-leading perturbations}
In \cite{CV}, sub-leading perturbation potentials are also discussed, for example, 
\[
W(z,t)= \frac{z^6}{6}- \frac{t  z^2}{2}
\]
which is simply rewritten as $t^{3/2} W_1(z^2/t^{1/2})/2$.
In the point of view of functional integrals over super-fields, 
it is argued that  ${\mathbf Q}_{\rm CFIV}$ must be twice of that for  $W_1(z)$.
Let us interpret this in terms of an ODE.

According to our working hypothesis, consider
\[
\psi(t)=\int_{\cal C} {\rm e}^{ W(z,t)} dz.
\]
It satisfies the $3^{\text{rd}}$ order ODE
\[
\frac{d^3}{dt^3} \psi(t) =\frac{1}{8}\bigl(\psi(t)+2 t \frac{d}{dt}\psi(t)\bigr)
\]
or in terms of a rescaled variable $x=t/(2)^{4/3}$,
\[
\frac{d^3}{dx^3} \phi(x) -4x\frac{d}{dx} \phi(x)-2\phi(x)=0
\]
where   $\phi(x)= \psi(2^{4/3} x)$.

It is well-known that ${\rm Ai}^2(x), {\rm Bi}^2(x)$ and ${\rm Ai}(x) {\rm Bi}(x)$  are 
solutions to this ODE.   
 Now we shift $x \rightarrow x-E$ and 
 take $\phi_j^2, \phi_{j+1}^2$ and $\phi_{j} \phi_{j+1}$ as
FSS in ${\cal S}_j$ where $\phi_j$ is defined in (\ref{yjsu2}).
Indeed,  it is easily checked that
\[
W[\phi_j^2,\phi_{j+1}^2,\phi_{j}\phi_{j+1}]=2(W[\phi_j,\phi_{j+1}])^2
\]
where $W$ denotes the Wronskian determinant.
Thus three functions are linearly independent.
The connection between ${\cal S}_0$ and ${\cal S}_{-1}$ is
easily solved.
It follows form the  three term relation of the Ai function that  
\[
 \phi_1^2 = \phi_0^2+\phi_{-1}^2 + 2 \phi_0 \phi_{-1} = \phi_0^2 +\phi_2^2+2 \phi_0 \phi_2.
\]
With nonzero angular momentum,  this may be modified as
\[
T \phi_1^2 = \phi_0^2+\phi_{-1}^2 + 2 \phi_0 \phi_{-1} = \phi_0^2 +\phi_2^2+2 \phi_0 \phi_2.
\]
The Stokes multiplier is thus squared $T=\tau_1^2$.
Since ${\mathbf Q}_{\rm CFIV}$  is essentially the logarithm of the Stokes multiplier,
this means  ${\mathbf Q}_{\rm CFIV}$  should be doubled,  in agreement with the argument in \cite{CV}.

More generally, for a perturbation potential
\[
W_{m}(z,t)= \frac{z^{2m}}{2m}-\frac{t}{2} z^2,
\]
the associated function 
\[
\psi_m (t)=\int_{\cal C} {\rm e}^{ W_m(z,t)} dz
\]
is found to satisfy a special case of so(m+1) ODE in the classification of \cite{DDMST},
\[
\frac{d^m}{dt^m} \psi_m(t)= (-1)^m  \sqrt{t}\frac{d}{dt} \sqrt{t} \psi_m(t).
\]
This is again consistent with the observation in \cite{CV}.

\subsection{Eigenvalues of  conserved quantities}

The formula for the vacuum expectation values of
conserved quantities $\mathbf{I}_{2n-1}$ in CFT based on $U_q(sl_2)$ symmetry  is  
conjectured in \cite{BLZ2}.
In the present framework, the result is translated to 
\[
\int {\rm e}^{m\theta}  \ln(1+\mathbf{Y}_k(\theta)) \frac{d\theta}{2\pi}
\propto 
\mathbf{I}_{m}
\]
for $k=1$ and $m=2n-1$.
The analytic evaluation of the above is difficult except for the $m=1$ case  to which
the dilogarithm technique can be applied.
This is due to the fact that the analytic expression of $Y_k$ is not available in general.

Although all  $\mathbf{I}_m$ become null  strictly at  $h=0$, we expect
 the left hand side brings conserved quantities order by order in $h$.
 Indeed,  the first order quantity in $h$  for $m=1$
 agrees with the CFIV index.
Let us assume that this conjecture is valid for arbitrary $k$ and $m$ and evaluate 
\[
\widetilde{\mathbf{I}}_m := \int {\rm e}^{m\theta} {\rm e}^{-A(\theta )}\frac{d\theta}{2\pi}.
\]
Since we have the explicit solution to ${\rm e}^{-A(\theta )}$, the evaluation of 
$\widetilde{\mathbf{I}}_m$ is immediate.
The actual calculation goes parallel to the one for the CFIV index.
Thanks to (\ref{Cecottiformula}), we find
\[
\widetilde{\mathbf{I}}_m= \frac{2^{2(m-1)} }{\pi} 
\frac{\Gamma(\frac{m}{2})^2 }{\Gamma(m) }
\frac{ \Gamma(\frac{m}{2}+\frac{1}{k+2}) \Gamma(\frac{m}{2}+1-\frac{1}{k+2})   } 
{\Gamma(\frac{1}{2}+\frac{1}{k+2}) \Gamma(\frac{1}{2}-\frac{1}{k+2})}.
\]

\section{Summary and conclusion}\label{sec:summary}

In this report,  we propose a  hypothesis that 
the combination of  the wave functions associated with $F$-term potentials 
yields  the Y functions of  the corresponding  massless TBA equations.
This has been successfully demonstrated for  nontrivial examples.
As applications, expectation values of conserved quantities, beyond the CFIV index, 
are conjectured by using the explicit solutions.

There remain obviously many questions  open.

The unexpected cancellation and the truncation of TBA equations, observed for
the SU(3) case, may be  generic for SU(n$\ge 3$), %
which needs a proof. 
Moreover, we need to understand the intrinsic reason why such a miraculous cancellation should occur.
 
The  super-potentials for the super-conformal theories, except for A type, contain multi-variables \cite{VafaWarner, Martinec}.
It  is not clear
how to extend the observation in  this report, especially how to define ``wave functions",   in  these cases.

The argument given in this communication  is restricted to
 the massless case, in which ODE equations 
help us to find solutions.
The original problem is concerned, however, with the generally massive quantum field theory: 
the TBA, or the CFIV index are developed for the analysis of such a case.
We note the recent progress in the ODE/IM correspondence 
towards  the massive deformation \cite{LukyanovZamolodchikovSh, DFNT, ItoLocke, AD14}. 
Hopefully it will help us to analyze the massive TBA through auxiliary linear problems associated to
integrable partial differential equations.

Ultimately, 
the reason why our hypothesis works well remains a mystery.
This is the  most serious problem at the present moment.
We hope to come back to this in the near future.

\subsection*{Acknowledgements} 
  The author would like to thank P.~Dorey, A.~Kl{\"u}mper and R.~Tateo for the critical reading of the manuscript.
  He also thanks for the warm hospitalities by the organizers of  ``infinite analysis 2014'' and ``integrable lattice models
  and quantum field theories" where a part of the present contents was presented.
  This work has been supported by  JSPS Grant-in-Aid for Scientific Research (C) No.  24540399.

\appendix
\section{Dressed Vacuum Form : SU(3) }\label{appendixDVFsu3}
We write down   (\ref{wronskiansu3}) in a way that the connection to integrable systems is obvious.
For this purpose, we introduce
\[
D^{(2)}(E,\mathbf{g})  =\Bigl( \prod_{0\le i<j \le 2} (g_j-g_i) \Bigr)
\bigl( D^{[0]}(E) D^{[1]}(E\Omega^3) q^{-g_1}
-D^{[1]}(E) D^{[0]}(E\Omega^3) q^{-g_0}  \bigr).
\]
By using the elementary formula on matrix determinants \cite{DDTsuN}, we can derive,
\[
\tau^{(1)}_1(E)=
\xi_0  \frac{D^{[0]}(E)} {D^{[0]}(E \Omega^3)} 
+
\xi_1 \frac{D^{[0]}(E\Omega^6)} {D^{[0]}(E \Omega^3)} \frac{D^{(2)}(E)} {D^{(2)}(E \Omega^3)} 
+
\xi_2 \frac{D^{(2)}(E\Omega^6)} {D^{(2)}(E \Omega^3)},
\]
where we set
\[
\xi_0=q^{g_0-1},\qquad 
\xi_1=q^{g_1-1}, \qquad
\xi_2=q^{g_2-1}.
\]
Further,  use $\Omega^3=q^{-1}$ and  write
\[
{\mathbf Q}(E)=D^{[0]}(E), \quad
\bar{{\mathbf Q}}(E)=D^{(2)}(Eq^{\frac{1}{2}}), \quad
T^{(1)}_1(E)= \tau^{(1)}_1(E q^{\frac{5}{4}}),
\]
then we obtain the expression,
\[
{\bf  T}^{(1)}_1(E)=
\xi_0  \frac{{\mathbf Q}(E q^{\frac{5}{4}})} {{\mathbf Q}(E q^{\frac{1}{4}})} 
+
 \xi_1 \frac{{\mathbf Q}(E q^{-\frac{3}{4}})} {{\mathbf Q}(E q^{\frac{1}{4}})}
 \frac{\bar{{\mathbf Q}}(E  q^{\frac{3}{4}})} {\bar{{\mathbf Q}}(E  q^{-\frac{1}{4}})} 
+
\xi_2 \frac{\bar{{\mathbf Q}}(Eq^{-\frac{5}{4}})} {\bar{{\mathbf Q}}(E q^{-\frac{1}{4}})}.
\]

This agrees with the known DVF for the  integral model  with $A^{(1)}_2$ symmetry.
Similarly, by setting 
$T^{(2)}_1(E)= \tau^{(2)}_1(E q^{\frac{7}{4}})$
one finds,
\[
{\bf  T}^{(2)}_1(E)=
\xi'_0 \frac{ \bar{{\mathbf Q}}  (E q^{\frac{5}{4}})} {\bar{{\mathbf Q}} (E q^{\frac{1}{4}})} 
+
\xi'_1
 \frac{\bar{{\mathbf Q}}(E  q^{-\frac{3}{4}})} {\bar{{\mathbf Q}}(E  q^{\frac{1}{4}})} 
 \frac{{\mathbf Q}(E q^{\frac{3}{4}})} {{\mathbf Q}(E q^{-\frac{1}{4}})}
+
\xi'_2 \frac{{\mathbf Q}(Eq^{-\frac{5}{4}})} {{\mathbf Q}(E q^{-\frac{1}{4}})}
\]
where
\[
\xi'_0=\xi_0 \xi_1 = (\xi_2)^{-1}, 
\qquad 
\xi'_1=(\xi_1)^{-1}, 
\qquad  
 \xi'_2=(\xi_0)^{-1}.
\]

\section{ Solutions to the $3^\text{rd}$  order ODE}  \label {appendixodesolution}
There are elementary functions that solve the $3^\text{rd}$ order equations, but are 
not widely known  in contrast to the $2^\text{nd}$  order case.
It is shown, however, in \cite{Braaksma} that the solution $\varphi(x)$ to  (\ref{ODEsu30})
with the desired asymptotic behavior (\ref{j:asym})  is given by
Meijer's G function\footnote{to meet (\ref{ODEsu30}), we rotate
$x \rightarrow {\rm e}^{i\pi/4}x$ .}
,
\[
\varphi(x) = c\, G^{3,0}_{0,3}
\left( { \phantom{a} - \phantom{b}
  \atop   0,\frac{1}{4},\frac{2}{4}} |  \frac{x^4}{4^3}  \right)
\]
for $0\le {\text{arg}}\,(x^4/4^3) \le  6\pi$.
The normalization constant  $c$ should be determined so as to be consistent with (\ref{j:asym}).
 To verify this, one requires several machineries.  
 
 In this appendix, we  take a formal but   simple approach to represent the solution $\varphi(x)$:
 we try to represent it with a familiar object, 
 a limit of the generalized hypergeometric series ${}_3 F_2$.
The following argument generalizes the one given for the Airy function 
in its relation to finite size lattice models \cite{DST}.

The  generalized hypergeometric series $_3 F_2(a_1, a_2, a_3; b_1, b_2| \zeta)$ satisfies
the following differential equation,
\begin{align}
&[\delta(\delta+b_1-1)(\delta+b_2-1) -z (\delta +a_1)  (\delta +a_2) (\delta +a_3)]{}_3F_2 =0, 
\label{gHG} \\
&\phantom{abc} \delta :=  \zeta \frac{d}{d \zeta}. \nonumber
\end{align}
In the following we  adopt abbreviations ${\bold a}:=(a_1, a_2, a_3), {\bold b}=(b_1, b_2)$
and $_3 F_2({\bold a};{\bold b}| \xi)$.

Consider  the limit
\[
 %
 \lim_{N \rightarrow \infty} \phantom{}_3 F_2 
(\,{{\bold a} \atop {\bold b}} \, |  \, \frac{x^L}{L^3 N(N^2-1)} ).
\]
It is easy to check that it satisfies (\ref{ODEsu30}) where $L=4, \sigma=-\frac{1}{4}$   and 
\[
{\bold a}=( -N, -N+ \sigma, -N +2\sigma), \qquad 
{\bf b} = (1+\sigma, 1+2\sigma) .
\]
The parameter $N$ corresponds to the lattice size for the  SU(2) case \cite{DST}.

The desired solution, $\varphi$,  however should satisfy  the asymptotic behavior (\ref{j:asym}).
In order to accomplish this,  we first keep $N$ a large but finite positive integer
and use the fact that 
two other  independent solutions to  (\ref{gHG})
are
\[
\zeta^{-\sigma} \phantom{}_3 F_2  ({\bold a}'; {\bold b}'  |\zeta)
\qquad \text{and} \qquad 
\zeta^{-2\sigma} \phantom{}_3 F_2  ({\bold a}''; {\bold b}''  | \zeta) 
\]
where 
${\bold a}'= {\bold a}-\sigma (1,1,1), {\bold b}'=(1-\sigma, 1+\sigma)$ and
${\bold a}''= {\bold a}-2\sigma (1,1,1), {\bold b}''=(1-\sigma, 1-2\sigma)$.

Let a subsidiary variable  $z^L=\zeta$ and 
a subsidiary function $\Psi_{N,\sigma}$ be
\[
\Psi_{N,\sigma}(x) =\frac{z^{L \sigma+1 }}{(1-z^L)^N } \,\phantom{}_3 F_2({\bold a}; {\bold b}| z^L).
\]
The ODE for  $\Psi_{N,\sigma}$ reads,
\begin{align}\label{odepsi}
 & z^3 \frac{d^3}{dz^3}     \Psi_{N,\sigma} 
 -L^2  \frac{3 N(N+1) z^L}{(z^L-1)^2} z \frac{d}{dz}  \Psi_{N,\sigma}  \nonumber \\
 & +L^3 \frac{N(N^2-1)(z^L+1) z^L}{(1-z^L)^3}   \Psi_{N,\sigma}
+3L^2  \frac{N(N+1) z^L}{(1-z^L)^2}   \Psi_{N,\sigma} =0,
%
%
\end{align}
where $\sigma L=-1$ is used .

We select  a solution which  vanishes  at the regular singular point  $z=1$.
Let us  consider the
linear combination of  three independent solutions 
\[
\Psi_{N,\sigma}
:=
\frac{1} {(1-z^L)^N} (\alpha_1 \psi_1+ \alpha_2 \psi_2+\alpha_3 \psi_3 )
\]
where 
\begin{align*}
\psi_1 &=  z^{L \sigma+1} \,{}_3 F_2({\bold a}; {\bold b}; z^L),    \\
\psi_2 &=  z \,{}_3 F_2({\bold a}'; {\bold b}';z^L),  \\
\psi_3 &=  z^{-L \sigma+1} \,{}_3 F_2({\bold a}''; {\bold b}'';z^L).  \\  
\end{align*}

The characteristic exponents at $z=1$ of  (\ref {odepsi}) are $-N, -N+1, 2N+2$, thus it suffices to 
require
\[
(\alpha_1 \psi_1+ \alpha_2 \psi_2+\alpha_3 \psi_3 )|_{z=1}=0 , \quad
\frac{d}{dz}(\alpha_1 \psi_1+ \alpha_2 \psi_2+\alpha_3 \psi_3 )|_{z=1}=0.
\]
These fix $\alpha_i$ (but for an overall factor),
\[
\alpha_1(N)=c\frac{W[\psi_2,\psi_3] }{W[\psi_1,\psi_2,\psi_3]},
\quad
\alpha_2(N)=-c\frac{W[\psi_1,\psi_3] }{W[\psi_1,\psi_2,\psi_3]},
\quad
\alpha_3(N)=c\frac{W[\psi_1,\psi_2] }{W[\psi_1,\psi_2,\psi_3]}
\]
where $W$ denotes the Wronskian determinant evaluated at $z=1$.

We assume the following limit exists,
\[
\alpha_i= \lim_{N\rightarrow \infty} \Bigl(\frac{1}{4^3 N (N^2-1)}  \Bigr)^{\frac{i-1}{4}} \alpha_i(N)
\]
although the actual evaluation of the limit may not be simple.

In the scaled variables $x$, the regular singular point is relocated at $x_0 = (4^3N(N^2-1))^{\frac{1}{4}}$.
By an  argument similar to the SU(2) case, we can show that $\Psi_{N,\sigma}$  is a decreasing function in 
$0<x<x_0$.  
In the ``scaling limit" $N \rightarrow \infty$, $x_0$ also goes to infinity.
Thus we conclude  that $\varphi = \lim_{N \rightarrow \infty}\Psi_{N,\sigma}  $ is the desired decaying function, 
 which is written as follows,
\[
\varphi(x)=\alpha_1\, \phantom{}_0 F_2(; \frac{3}{4},\frac{1}{2}|\frac{x^4}{4^3})
+\alpha_2  x\, \phantom{}_0 F_2(; \frac{3}{4},\frac{5}{4}|\frac{x^4}{4^3})
+ \alpha_3  x^2\,  \phantom{}_0 F_2(; \frac{5}{4},\frac{3}{2}|\frac{x^4}{4^3}).
\]

\section{The cancellation of terms }\label{appendixCancellation}

Using the expansions (\ref {Texpansionsu3}) into definitions of  (\ref{YTsu3}),
we obtain the expressions of $y^{(a)}_{j,m}(\theta)$ in terms of $t^{(a')}_{j',m'}(\theta)$.
Especially we are interested in 
$y^{(1)}_{2,2}(\theta)+ y^{(2)}_{2,3}(\theta)$ and  $y^{(1)}_{2,3}(\theta)+ y^{(2)}_{2,2}(\theta)$.
Since ${\bf T}^{(a)}_m$ satisfies the T-system,  $t^{(a')}_{j',m'}(\theta)$ are not necessarily independent:
there are some relations, e.g.,
\begin{align*}
t^{(1)}_{1,3}(\theta)&=- t^{(2)}_{1,2}(\theta), \qquad     t^{(2)}_{1,3}(\theta)=- t^{(1)}_{1,2}(\theta), \\
t^{(1)}_{1,4}(\theta)&=- t^{(2)}_{1,1}(\theta), \qquad     t^{(2)}_{1,4}(\theta)=- t^{(1)}_{1,1}(\theta), \\
t^{(1)}_{2,3}(\theta)&=t^{(1)}_{1,2}(\theta+i\frac{\pi}{3}) +t^{(1)}_{1,2}(\theta-i\frac{\pi}{3}) +t^{(1)}_{1,1}(\theta) t^{(2)}_{1,2}(\theta) 
-t^{(2)}_{2,2}(\theta),\\
t^{(2)}_{2,3}(\theta)&=t^{(2)}_{1,2}(\theta+i\frac{\pi}{3}) +t^{(2)}_{1,2}(\theta-i\frac{\pi}{3}) +t^{(2)}_{1,1}(\theta) t^{(1)}_{1,2}(\theta) 
-t^{(1)}_{2,2}(\theta)\\
\end{align*}
and so on.
Using these relations, we find  relatively simple expressions,
\begin{align*}
y^{(1)}_{2,3}(\theta)+ y^{(2)}_{2,2}(\theta) &= (t^{(2)}_{1,1}(\theta))^2 -
 (  \frac{t^{(2)}_{1,2}(\theta+i\frac{\pi}{3}) t^{(2)}_{1,2}(\theta-i\frac{\pi}{3})}{ t^{(1)}_{1,2}(\theta)     }       )^2 
 -t^{(1)}_{2,4}(\theta)-t^{(2)}_{2,1}(\theta), \\
 y^{(1)}_{2,2}(\theta)+y^{(2)}_{2,3}(\theta) &= (t^{(1)}_{1,1}(\theta))^2 -
 (  \frac{t^{(1)}_{1,2}(\theta+i\frac{\pi}{3}) t^{(1)}_{1,2}(\theta-i\frac{\pi}{3})}{ t^{(2)}_{1,2}(\theta)     }       )^2 
-t^{(1)}_{2,1}(\theta) -t^{(2)}_{2,4}(\theta).
\end{align*}
They however still show that we need $O(h^2)$ terms $t^{(a)}_{2,m}$  in the expansions of ${\bf T}^{(a)}_m$.
Note that 
\begin{equation}\label{y212y112}
y^{(2)}_{1,2}(\theta) =\frac{t^{(2)}_{1,2}(\theta+i\frac{\pi}{3}) t^{(2)}_{1,2}(\theta-i\frac{\pi}{3})}{ t^{(1)}_{1,2}(\theta)     } 
\quad \text{and} \quad 
y^{(1)}_{1,2}(\theta) =\frac{t^{(1)}_{1,2}(\theta+i\frac{\pi}{3}) t^{(1)}_{1,2}(\theta-i\frac{\pi}{3})}{ t^{(2)}_{1,2}(\theta)     } 
\end{equation}
also appear in the last term of   (\ref{etaeq1}).

The key idea is to employ the expansions of $D^{[i]}(E,{\bf g} )$,
\begin{equation*}
D^{[i]}(E,{\bf g} ) =d^{[i]}_{0}(E)+ h  d^{[i]}_{1} (E)+ h^2  d^{[i]}_{2} (E) +O(h^3), \quad (i=0,1,2).
\end{equation*}

Although $d^{[i]}_{0}(E)$ is denoted by $Q^{[i]}(E)$ in the main body, we use   $d^{[i]}_{0}$ in the appendix for uniformity.
Substituting this into (\ref{qWronskiansu3}), taking account of  the shift in (\ref{Ttaurelation}) and (\ref{EthetasuN}),
we are able to derive the expressions of $t^{(a)}_{j, m}$ in terms of 
 $d^{[i]}_{j}$.
 The point is that the latter has a smaller number of 
 independent elements (for fixed order of expansions in $h$) and 
 a smaller number of  relations.
 
 The obvious constraint  is $\tau^{(1)}_0(E,{\bf g})=1$. We write it as
 \[
 \tau^{(1)}_0(E,{\bf g})=1= \tau^{(1)}_{0,0} (E)+ h \tau^{(1)}_{1,0} (E)+h^2 \tau^{(1)}_{2,0} (E)+O(h^2)
 \]
 or 
 \[
 \tau^{(1)}_{0,0} (E)=1,  \quad  \tau^{(1)}_{1,0} (E)=0,  \quad \tau^{(1)}_{2,0} (E)=0 \cdots.
 \]
 
 The constraints above are written in terms of sums of triple products of   $d^{[i]}_{j}$.
  By comparing the expressions in terms of  $d^{[i]}_{j}$,
  we find
  \begin{align*}
  &t^{(2)}_{1,1}(\theta) +
 \frac{t^{(2)}_{1,2}(\theta+i\frac{\pi}{3}) t^{(2)}_{1,2}(\theta-i\frac{\pi}{3})}{ t^{(1)}_{1,2}(\theta)     }     
 =-i   \tau^{(1)}_{0,0} (-E {\rm e}^{\frac{3}{8}\pi}) =-i ,  \\
&t^{(1)}_{1,1}(\theta) +
 \frac{t^{(1)}_{1,2}(\theta+i\frac{\pi}{3}) t^{(1)}_{1,2}(\theta-i\frac{\pi}{3})}{ t^{(2)}_{1,2}(\theta)     }     
 = i   \tau^{(1)}_{0,0} (-E {\rm e}^{\frac{5}{8}\pi}) =i .
  \end{align*}
There are also unexpected relations,
  \begin{align*}
t^{(2)}_{1,1}(\theta) &-
\frac{t^{(2)}_{1,2}(\theta+i\frac{\pi}{3}) t^{(2)}_{1,2}(\theta-i\frac{\pi}{3})} { t^{(1)}_{1,2}(\theta) }   +i   \tau^{(1)}_{0,0} (-E {\rm e}^{\frac{3}{8}\pi}) \\ 
&=t^{(2)}_{1,1}(\theta) -
\frac{t^{(2)}_{1,2}(\theta+i\frac{\pi}{3}) t^{(2)}_{1,2}(\theta-i\frac{\pi}{3})} { t^{(1)}_{1,2}(\theta) }   +i   \\ 
 &=6 (- d^{[0]}_0( E \omega^7) d^{[1]}_0( E \omega^3 )
 +  i  d^{[0]}_0( E \omega^{3}) d^{[1]}_0( E \omega^{7})) d^{[2]}_0(- E \omega^{3}),
 \\
t^{(1)}_{1,1}(\theta) &-
 \frac{t^{(1)}_{1,2}(\theta+i\frac{\pi}{3}) t^{(1)}_{1,2}(\theta-i\frac{\pi}{3}) }   { t^{(2)}_{1,2}(\theta)  }    -i  \tau^{(1)}_{0,0} (-E {\rm e}^{\frac{5}{8}\pi})\\
 &=t^{(1)}_{1,1}(\theta) -
 \frac{t^{(1)}_{1,2}(\theta+i\frac{\pi}{3}) t^{(1)}_{1,2}(\theta-i\frac{\pi}{3}) }   { t^{(2)}_{1,2}(\theta)  }    -i  \\
&=6    (    i d^{[0]}_0( -E \omega^5) d^{[1]}_0( -E \omega ) 
 + d^{[0]}_0( -E \omega) d^{[1]}_0( -E \omega^5 ) )d^{[2]}_0( E \omega^5)
  \end{align*}
where $\omega={\rm e}^{\frac{\pi}{8}i}$.
We also find
\begin{align*}
%
%
& t^{(2)}_{2,1}(\theta)+ t^{(1)}_{2,4}(\theta) +  1\\
&= t^{(2)}_{2,1}(\theta)+ t^{(1)}_{2,4}(\theta) +  \tau^{(1)}_{0,0} (-E \omega^3)-
2 \tau^{(1)}_{2,0} (-E \omega^3) \\
&=-3   \bigl ( d^{[0]}_0( E \omega^3) d^{[1]}_0( E \omega^7 ) 
+ i d^{[0]}_0( E \omega^7) d^{[1]}_0( E \omega^3) \bigr )d^{[2]}_0( -E \omega^3), \\
%
&t^{(1)}_{2,1}(\theta)+ t^{(2)}_{2,4}(\theta) + 1 \\
&=t^{(1)}_{2,1}(\theta)+ t^{(2)}_{2,4}(\theta) +  \tau^{(1)}_{0,0} (-E  \omega^5)
-2 \tau^{(1)}_{2,0} (-E \omega^5)\\
&=3   \bigl (  d^{[0]}_0( -E \omega^5) d^{[1]}_0( -E \omega ) 
- i d^{[0]}_0( -E \omega) d^{[1]}_0( -E \omega^5 ) \bigr )d^{[2]}_0( E \omega^5).
\end{align*}

Using these results, the sums of our interest simplify considerably,
\begin{align*}
y^{(1)}_{2,3}(\theta)+ y^{(2)}_{2,2}(\theta) &= 
9(d^{[0]}_0(E \omega^3) d^{[1]}_0(E \omega^7) + i d^{[0]}_0(E \omega^7) d^{[1]}_0(E \omega^3)) d^{[2]}_0(-E \omega^3), 
 \\
y^{(1)}_{2,2}(\theta)+y^{(2)}_{2,3}(\theta) &= 
9(i d^{[0]}_0(-E \omega) d^{[1]}_0(-E \omega^5) -d^{[0]}_0(-E \omega^5) d^{[1]}_0(-E \omega)) d^{[2]}_0(E \omega^5) .
\end{align*}

The right hand sides of (\ref{y212y112}) are also derived from the above results,
\begin{align}\label{solB0B0b}
y^{(2)}_{1,2}(\theta) &=3( d^{[0]}_0(E \omega^7) d^{[1]}_0(E \omega^3) -i d^{[0]}_0(E \omega^3) d^{[1]}_0(E \omega^7))
d^{[2]}_0(-E\omega^3),  \nonumber \\
y^{(1)}_{1,2}(\theta) &=-3( d^{[0]}_0(-E \omega) d^{[1]}_0(-E \omega^5)+ i d^{[0]}_0(-E \omega^5) d^{[1]}_0(-E \omega))
d^{[2]}_0(E\omega^5).
\end{align}

Thereby, we conclude the validity of  (\ref{simplification}).
%


\clearpage
\begin{thebibliography}{10}



\bibitem{Takahashibook}
M.~Takahashi, 
``Thermodynamics of One-Dimensional Solvable Models",
Cambridge University Press, (1999).

\bibitem{AlZPotts}
Al.~B.~Zamolodchikov,
``Thermodynamic Bethe ansatz in relativistic models: Scaling 3-state Potts and Lee-Yang models", 
Nucl.~Phys.~{\bf B 342} (1990) 695-720.

\bibitem{DTP}
P.~Dorey, A.~Pocklington and R.~Tateo,
``Integrable aspects of the scaling q-state Potts models I: bound states and bootstrap closure",
Nucl.~Phys.~{\bf B 661} (2003) 425-463.


\bibitem{Gaudin}
M.~Gaudin,
``Thermodynamics of the Heisenberg-Ising Ring for $\Delta \ge 1$",
Phys. Rev. Lett.  {\bf26} (1971)  1301.

\bibitem{TakahahiSuzuki}
M.~Takahashi and M.~Suzuki,
``One-Dimensional Anisotropic Heisenberg Model at Finite Temperatures",
Prog.~Theoret.~Phys.{\bf 48} (1972) 2187-2209.

\bibitem{KlumperRSOS}
A.~Kl{\"u}mper,
``Free energy and correlation lengths of quantum chains related to restricted solid-on-solid lattice models",
Ann.~Phys.{\bf 504} (1992) 540-553.

\bibitem{JKSfusion} 
G.~J{\"u}ttner, A.~Kl{\"u}mper and J.~Suzuki, 
``From fusion hierarchy to excited state TBA",
Nucl.~Phys.~{\bf B 512} (1998) 581-600.

\bibitem{DRTW}
P.~Dorey, I.~Runkel, R.~Tateo and  G.~Watts,
``g-function flow in perturbed boundary conformal field theories
",  Nucl.~Phys. {\bf B578} (2000) 85-122

\bibitem{TsvelikWiegmann}
A.~M.~Tsvelick and P.~B.~Wiegmann,
`` Exact results in the theory of magnetic alloys",
Adv. in Phys. {\bf 32} (1983) 453-713.

\bibitem{Fendley99} 
P.~Fendley, ``Airy functions in thermodynamic Bethe ansatz",
Letters in Math. Phys. {\bf 49} (1999) 229-233.

\bibitem{MTW}
B.~M.~McCoy, C.~ Tracy and T.~T.~Wu,
``Painlev{\'e} functions of the third kind",
J.~Math.~Phys. {\bf 18} (1977) 1058.

\bibitem{AlZ}
Al.~B.~Zamolodchikov,
``Painlev{\'e} III and 2D polymers",
Nucl.~Phys.~{\bf B 432} (1994) 427-456.

\bibitem{TW}
C.~ A.~Tracy, H.~Widom,
``Proofs of two conjectures related to the
thermodynamic Bethe ansatz",
Comm.~Math.~Phys. {\bf 179} (1996) 667-680.

\bibitem{FI}
P.~Fendley and K.~Intriligator,
``Scattering and thermodynamics of fractionally-charged supersymmetric solitons",
Nucl.~Phys.~{\bf B 372} (1992) 533-558.




\bibitem{CFIV}
S.~Cecotti, P.~Fendley, K.~Intriligator and C.~Vafa
``A new supersymmetric index",
Nucl.~Phys.~{\bf B 386} (1992) 405-452.


\bibitem{DDT}
P.~Dorey, T.~C.~Dunning, R.~Tateo
``The ODE/IM Correspondence",
J.~Phys. {\bf A40} (2007) R205.


\bibitem{VafaWarner}
C.~Vafa and N.~P.~Warner
``Catastrophes and the Classification of Conformal Theories ",
Phys.~Lett. {\bf B218} (1989) 51. 

\bibitem{Martinec} 
E.~J.~Martinec,
``Algebraic Geometry and Effective Lagrangians ",
Phys.~Lett. {\bf B217} (1989) 431.

\bibitem{CV} 
S.~Cecotti  and C.~Vafa,
``Topological-anti-topological fusion",
Nucl.~Phys.~{\bf B 367} (1991) 359-461.


\bibitem{DT98}
P.~Dorey, R.~Tateo,
``Anharmonic oscillators, the thermodynamic Bethe ansatz and nonlinear integral equations",
J.~Phys. {\bf A32} (1999) L419.


\bibitem{DT99} 
P.~Dorey, R.~Tateo,
``On the relation between Stokes multipliers and the T-Q systems of conformal field theory",
Nucl.~Phys.~{\bf B 563} (1999) 573-602.

\bibitem{Baxterbook}
R.~Baxter,
``Exactly solved models in statistical mechanics", Academic Press  (1982) .


\bibitem{BLZODE}
V.~V.~Bazhanov, S.~L.~Lukyanov and A.~B.~Zamolodchikov,
``Spectral determinants for Schr{\"o}dinger equation and Q-operators of conformal feld theory",
J.~Stat.~Phys. {\bf 102}(2001) 567-576.



\bibitem{BLZ2}
V.~V.~Bazhanov, S.~L.~Lukyanov and A.~B.~Zamolodchikov,
``Integrable Structure of Conformal Field Theory II. Q-operator and DDV equation",
Comm.~Math.~Phys. {\bf 190}(1997) 247-278.


\bibitem{BLZ1}
V.~V.~Bazhanov, S.~L.~Lukyanov and A.~B.~Zamolodchikov,
``Integrable structure of conformal field theory, quantum KdV theory and Thermodynamic Bethe Ansatz",
Comm.~Math.~Phys. {\bf 177}(1996) 381-398.

\bibitem{Lukyanovup}
S.~L.~Lukyanov, 
``Notes on parafermionic QFT's with boundary interaction",
Nucl.~Phys.~{\bf B 784} (2007) 151-201.



\bibitem{DDMST}
P.~Dorey, T.~C.~Dunning, D.~Masoero, J.~Suzuki and R.~Tateo,
``Pseudo-differential equations, and the Bethe ansatz for the classical Lie algebras",
Nucl.~Phys.~{\bf B 772} (2007) 249-289.

\bibitem{KluemperPearce}
A.~Kl{\"u}mper and P.~Pearce,
``Conformal weights of RSOS lattice models and their fusion hierarchies",
Physica  {\bf A 183} (1992) 304.

\bibitem{KNS}
A.~Kuniba, T.~Nakanishi and J.~Suzuki,
``Functional relations in  solvable lattice models I: functional relations and representation theory ",
Int.~J.~Mod.~Phys., {\bf A9} (1994)  5215.

\bibitem{ZamolodchikovY}
Al.~B.~Zamolodchikov,
``On the thermodynamic Bethe ansatz equations for reflectionless ADE scattering theories", 
Phys.~Lett.~{\bf B253} (1991) 391-394.

\bibitem{AlZamolodchikovmassless}
Al.~B.~Zamolodchikov,
`` From tricritical Ising to critical Ising by thermodynamic Bethe ansatz ",
Nucl.~Phys.~{\bf B 358} (1991) 524- 546.

\bibitem{DTsu3}
P.~Dorey and  R.~Tateo,
``Differential equations and integrable models: the SU(3) case",
Nucl.~Phys.~{\bf B 571} (2001) 583-606.



\bibitem{BHK}
V.~V.~Bazhanov, A.~N.~Hibbard and  S.~M.~Khoroshkin,
``Integrable structure of  $W_3$ Conformal Field Theory, Quantum Boussinesq Theory and Boundary Affine Toda Theory",
Nucl.~Phys.~{\bf B 622} (2002) 475-547.


\bibitem{DDTsuN}
P.~Dorey, T.~C.~Dunning and R.~Tateo,
``Differential equations for general SU(n) Bethe ansatz systems",
J.~Phys. {\bf A33} (2000) 8427-8441.

\bibitem{JSsuN}
J.~ Suzuki,
``Functional relations in Stokes multipliers and solvable models related to $U_q(A^{(1)}_n)$",
J.~Phys. {\bf A33} (2000) 3507-3521.

\bibitem{DamerauKluemper}
J.~Damerau,
``Nonlinear integral equations for the thermodynamics of integrable quantum chains", 
(Thesis, Wuppertal University  2008)


\bibitem{Braaksma}
B.~L.~J.~Braaksma,
``Asymptotic Analysis of Differential Equation of Turrittin",
SIAM J. Math. Anal. {\bf 2} (1971) 1-16.


\bibitem{DST}
P.~Dorey, J.~Suzuki and  R.~Tateo,
``Finite lattice Bethe ansatz systems and the Heun equation",
J.~Phys. {\bf A37} (2004) 2047.


\bibitem{Cecottietal}
S.~Cecotti, L.~Girardello and A.~Pasquinucci,
``Non-perturbative aspects and exact results
for ${\cal N}=2$ Landau-Ginzburg models",
Nucl.~Phys.~{\bf B 328} (1989) 701-722.

\bibitem{LukyanovZamolodchikovSh}
S.~L.~Lukyanov, A.~B.~Zamolodchikov,
``Quantum sine(h)-Gordon model and classical integrable equations",
JHEP {\bf 07} (2010)008.

\bibitem{DFNT}
P.~Dorey,  S.~Faldella, S.~Negro and R.~Tateo,
``The Bethe ansatz and the Tzitz{\'e}ica-Bullough-Dodd equation",
Phil.~Trans.~R.~Soc. {\bf A 371} 20120052.

\bibitem{ItoLocke}
K.~Ito and C.~Locke,
``ODE/IM correspondence and modified affine Toda field equations",
Nucl.~Phys. {\bf B885} (2014) 600-619(arXiv:1312.6759).

\bibitem{AD14}
	P.~Adamopoulou and C.~Dunning, 
``Bethe Ansatz equations for the classical $A^{(1)}_n$ affine Toda field theories",
	J.~Phys. {\bf A47 }(2014) 205205  (arXiv:1401.1187).


	


\end{thebibliography}
\end{document}